\documentclass[twocolumn,amsmath,amssymb]{revtex4}

\usepackage{multirow}
\usepackage{graphicx}
\usepackage{subfigure}
\usepackage{color}

\usepackage{dcolumn}
\usepackage{bm}
\usepackage{array}
\usepackage{color}

\usepackage{setspace}
\usepackage{enumitem}

\begin{document}

\title{
 Optimal Disruption of Complex Networks
}
\author{
  Jin-Hua Zhao$^{1}$ and Hai-Jun Zhou$^{1,2,3}$\footnote{Corresponding
    author. Email: {\tt zhouhj@itp.ac.cn}}
}

\affiliation{
  $^1$Key Laboratory of Theoretical Physics, Institute of
  Theoretical Physics, Chinese Academy of Sciences,
  Zhong-Guan-Cun East Road 55,
  Beijing 100190, China \\
  $^2$School of Physical Sciences, University of Chinese Academy of Sciences,
  Beijing 100049, China \\
  $^3$Synergetic Innovation Center for Quantum Effects and Applications
  (SICQEA), Hunan Normal University, Changsha 410081, China
}
\date{May 31, 2016}

\begin{abstract}
  The collection of \textsl{all} the strongly connected components in a directed graph,
  among each cluster of which any node has a path to another node,
  is a typical example of the intertwining structure and dynamics in complex networks,
  as its relative size indicates network cohesion
  and it also composes of all the feedback cycles in the network.
  Here we consider finding an optimal strategy with minimal effort in removal arcs
  (for example, deactivation of directed interactions)
  to fragment \textsl{all} the strongly connected components
  into tree structure with no effect from feedback mechanism.
  We map the optimal network disruption problem
  to the minimal feedback arc set problem,
  a non-deterministically polynomial hard combinatorial optimization problem in graph theory.
  We solve the problem with statistical physical methods from spin glass theory,
  resulting in a simple numerical method to extract sub-optimal disruption arc sets
  with significantly better results than a local heuristic method
  and a simulated annealing method
  both in random and real networks.
  Our results has various implications
  in controlling and manipulation of real interacted systems.
\end{abstract}

\maketitle




 
\section*{Introduction}

(\emph{The preprint is a working paper. It will be further revised. 
  Comments are welcome.})

In complex systems modeling as networks
\cite{
Boccaleti.Latora.Moreno.Chavez.Hwang-PhysRep-2006},
the constituents are considered as nodes or vertices,
and interactions are considered as links or arcs.
There are many examples of the embedded structure in networks
showing a dynamical significance.
The intertwined complexity of the structural topology and the dynamical behaviors
is especially typical in directed networks.
From the structural side,
the strongly connected components (SCC) of the directed networks
\cite{
Dorogovtsev.Mendes.Samukhin-PRE-2001},
in which any two nodes has certain path following consecutive and non-intersecting directed arcs to each other,
is a well-known indicator as the cohesion of the networks.
From the dynamical side,
in many complex systems with directed interactions,
the delicate control mechanisms to maintain stable functioning against external perturbations
(such as circadian rhythm in animals and plants)
or some irreversible decision-making processes
(such as apoptosis of cells and cancer growth in human tissues)
are results of architecture of feedback loops
 \cite{
Ingalls-2013,
Aguda.etal-PNAS-2008},
and the dynamics of an interaction topology without feedback loops are relatively easy to be driven
\cite{
Fiedler.Mochizuki.Kurosawa.Saito-JDynDiffEquat-2013a,
Mochizuki.Fiedler.Kurosawa.Saito-JTheorBiol-2013b}.
Our starting point for the paper is a simple truth
that \textsl{all} the SCCs are simply the collection
of all the loops or cycles in the graphs or networks.
An intuitive question naturally arises:
how we can disrupt \textsl{all} the SCCs,
correspondingly \textsl{all} the loops,
by the removal of a minimal number of nodes or arcs
thus there are only tree-like structures left with trivially dynamical significance?

The optimal network disruption problem
is closely related to the study on network resilience and robustness
\cite{
Albert.Jeong.Barabasi-Nature-2000,
Cohen.Erez.benAvraham.Havlin-PRL-2000}
since the inception of the network research
and the optimal percolation
\cite{
Morone.Makse-Nature-2015}
and network attack problem
\cite{
Mugisha.Zhou-arxiv-2016}
yet distances itself from them
as it provides an optimization perspective on the destruction protocol of directed networks,
a more realistic model of description of interactions in real interacted systems.
The arc direction in networks
leads to much different handling methods with previous research on the optimization problem:
we focus on the SCCs
rather than weakly connected components
(the largest component of the nodes
while every two nodes have certain directed paths between them)
\cite{
Molloy.Reed-RandStruAlgo-1995}
as the former has a more involved significance in the dynamics
apart from the structure;
we consider the removal of \textsl{all} the SCCs
rather than the \textsl{giant} strongly connected component (GSCC),
thus result in a principled method without
the problem of thresholding in finite graphs (the definition of how 'microscopic' is 'microscopic')
as present in the case of undirected networks considering the giant connected component.

\section*{Equivalence of strongly connected components and loops or cycles}

\begin{figure*}
\begin{center}
 \includegraphics[width = 0.65 \linewidth]{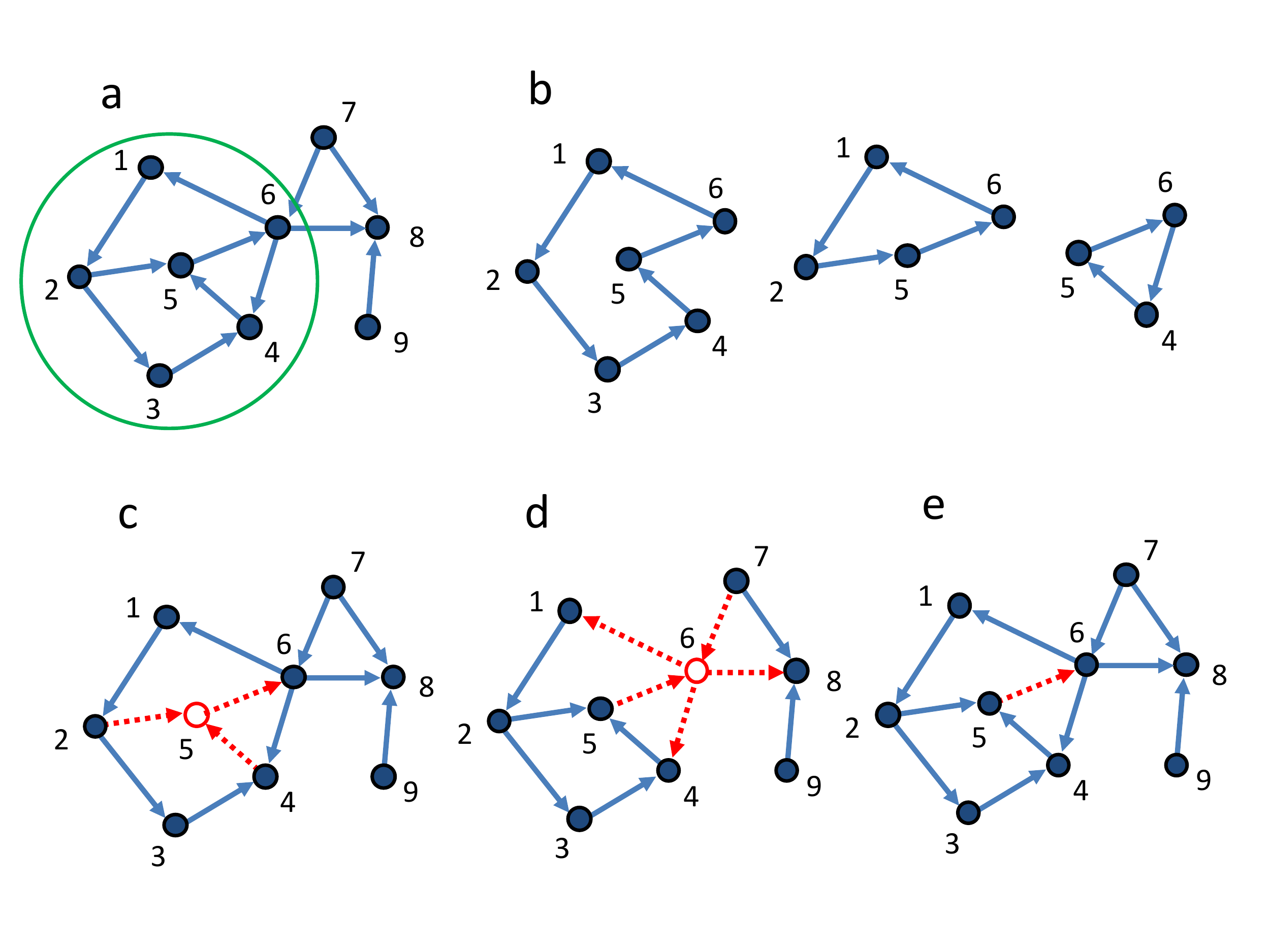}
\end{center}
\caption{
 \label{fig:cycles}
\textbf{Optimal network disruption of a small directed graph.}
\textbf{a,}
The small directed graph has $9$ vertices and $12$ directed arcs.
For simplicity, the weight on each node and arc is set uniform.
There is only one single SCC as
the vertex set of $\{1, 2, 3, 4, 5, 6\}$ and all the arcs among them, marked in the green circle.
\textbf{b,}
There are $3$ cycles in the graph as $(1, 2, 3, 4, 5, 6, 1)$, $(1, 2, 5, 6, 1)$, and $(4, 5, 6, 4)$,
which in all forms the single SCC.
\textbf{c} and \textbf{d,}
The disruption procedure by removing vertices
(along with all their adjacent arcs)
is considered.
Removing the vertex $5$ or $6$
(marked in red empty cycles, and their adjacent arcs to be removed along are marked in a dashed form)
both leads to the removal of all the cycles and correspondingly the disruption of the single SCC.
Thus the minimal disruption vertex set and the minimal feedback vertex set of the network are both $\{5\}$ or $\{6\}$ with size $1$.
\textbf{e,}
The disruption procedure by removing arcs is considered.
The removing of the arc $(5, 6)$ (marked in dashed form)
also results in the removal of all the cycles and the single SCC.
Thus the minimal disruption arc set and the minimal feedback arc set of the network are both $\{(5, 6)\}$ with size $1$.
}
\end{figure*}

As a starting point of the problem in this paper,
we present the equivalence of \textsl{all} the strongly connected components
and \textsl{all} the loops or cycles in the same network instance.
A directed network $D = \{V, A\}$ has
a vertex set $V = \{1, 2, ..., N\}$ ($|V| = N$) and
an arc set $A = V \times V$ ($|A| = M$)
with the arc density $\alpha \equiv M / N$.
An arc $(i, j)$ is an ordered pair of nodes with the predecessor $i$ pointing to the successor $j$.
A path between two nodes $l$ and $n$ is a consecutive sequence of
arcs such as $(l, m_1), (m_1, m_2), (m_2, m_3), ... (m_k, n)$
(which afterwards can be easily denoted as $(l, m_1, m_2, ..., m_k, n)$)
with $k + 1$ arcs which are non-intersecting ($l \neq m_1 \neq m_2 ... \neq n$).
The SCCs of network $D$
are those clusters of nodes
in each of which any two nodes $i$ and $j$ have a certain path to each other:
if node $i$ has a path to node $j$ as $(i, i_1, i_2, .. , i_k, j)$,
and node $j$ has a path to node $i$ as $(j, j_1, j_2, ..., j_l, i)$,
then these two combined paths form a cycle or loop as $(i, i_1, i_2, ..., i_k, j, j_1, j_2, ..., j_l, i)$,
or a consecutive and non-intersecting closed sequence of arcs.
Thus \textsl{all} the SCCs for a directed network
is simply the aggregate structure of the nodes and arcs in \textsl{all} of its cycles.
Two methods
(a leaf removal method and Tarjan's method
\cite{
Tarjan-SIAMJourOnComp-1972})
to decompose a directed network into SCCs and
a reproduction of the mean-field theory for the GSCC
can be found in Supplementary Information section I.

\section*{From the optimal network disruption to the minimal feedback vertex/arc set problem}

The destruction of \textsl{all} the SCCs in a directed system
corresponds to the removal of all the cycles in the same system.
Two simple procedures to remove all the cycles can be considered:
the removal of vertices (along with all their adjacent arcs),
correspondingly the finding of a disruption vertex set (DVS) is just the feedback vertex set problem (FVS)
and the finding of a minimal disruption vertex set (MDVS) is just the minimal feedback vertex set problem (MFVS);
the removal of arcs,
correspondingly the finding of a disruption arc set (DAS) is just the feedback arc set problem (FAS),
and the finding of a minimal disruption arc set (MDAS) is just the minimal feedback arc set problem (MFAS)
\cite{
Du.Pardalos-HandbookComOpti-1999}.
In a general case,
each vertex or arc has a predefined positive weight to account their cost in the removal process,
and the total cost to minimize can be further defined as the sum of weights
of the disruption vertex or arc set.
Both MFVS and MFAS are non-deterministically polynomial-hard (NP-hard) problems
\cite{
Garey.Johnson-1979}
which in the worst case have an exact algorithm
with an exponential computation time of the problem size
(such as the size of vertices of the graphs on which the problem is defined).
The MFAS can be considered as the MFVS
on a transformed directed graphs,
and can also be considered as a minimum dominating set problem (MDS)
\cite{
Haynes.Hedetniemi.Slater-1998,
Zhao.Zhou-JStatPhys-2015}
of a bipartite graph of vertices and arcs
(Supplementary Information section II).
An example of the optimal network disruption problem on a small directed graph
is in Fig.\ref{fig:cycles}.

Here we consider the MDAS/MFAS problem on directed graphs,
since the removal of arcs is a more controlled way of local perturbation of network structure.
(Afterwards, DAS and FAS are used interchangeably.)
Optimization problems
usually concerns finding the minimal energy among the configurations
which satisfy all the constraints defined on the graph structure.
Generally speaking, typical types of the constraints are
local constraints and global constraints.
Local constraints
are usually formulated on arcs or vertices with their nearest neighbors,
whose local structures makes them relatively direct for
an adoption of the cavity method from spin glass theory
\cite{
Mezard.Montanari-2009}.
Yet, the evaluation of each global constraint
needs considering a non-localized structure
or even all the nodes or links of the graph,
which brings along severe difficulty in introducing statistical mechanical approaches.
That is why the optimal network disruption problem
needs more involved methods to tackle than the above-mentioned hard problems.
Typical optimizations with global constraints are
the prize-collecting Steiner tree problem
\cite{
Bayati.etal-PRL-2008},
the feedback vertex set problem on undirected graphs
\cite{
Zhou-EPJB-2013},
which devise tailored auxiliary methods to transform the global constraints into a localized form
thus make possible the application of the statistical mechanical method.
Here we follow the same logic to apply a representation to render the global constraints on loops localized
before we apply the cavity method.

\section*{Height representation}

For a directed graph $D = \{V, A\}$ with a vertex set $V$ and an arc set $A$,
each arc $(i, j)$ has a predefined positive weight $w_{ij}$
as the cost of removing the arc.
If we only consider the size of disruption arc set, the weight can be set uniform.
On each node $i$ we assign a positive integer $h_i \in [0, H - 1]$ as its height,
while $H$ is the maximal height chosen for $D$.
Thus we have a height configuration as $\underline{h} \equiv \{h_1, h_2, ..., h_N\}$.
To account the direction on each arc $(i, j)$,
we define a height relation as $h_i > h_j$,
which is much like a potential decreasing along the arc direction.
Yet the existence of cycles
leads to at least one arcs violating the height relation in each cycle.
For example, in a small cycle with only three arcs $(l, m, n, l)$,
we cannot satisfy such a height configuration as $h_l > h_m > h_n > h_l$.
The cycles
thus bring a nontrivial effect on the assignment of heights on a directed graph
simply based on the height relation and the direction of each arc.
Removal of all the arcs with end-nodes violating the height relation
leaves a height configuration with satisfied height relation on all the residual arcs,
correspondingly an acyclic directed graph.
Thus all the arcs violating the height relation
constitute a FAS $\Gamma$.
To be quantitative,
for any directed arc $(i, j) \in A$,
a binary state $s_{ij} = \{0, 1\}$
is defined as $(i, j)$ being in a FAS $\Gamma$ ($s_{ij} = 1$) or not ($s_{ij} = 0$).
Then for the ease of discussion, on any arc $(i, j)$ with $h_i, h_j \in [0, H - 1]$,
a compact form of the height constraint can be defined as

\begin{equation}
\label{eq:Cij}
C_{ij}(h_i, h_j)
 = \theta (h_i - h_j) \delta _{s_{ij}}^{0}
 + [1 - \theta (h_i - h_j)] [1 - \delta _{s_{ij}}^{0}],
\end{equation}
where $\theta(x)$ is the Heaviside function
as $\theta (x) = 1$ when $x > 0$ and $0$ when $x \le 0$.
For any directed arc $(i, j)$,
$C_{ij}(h_i, h_j)$ is $1$
only if
$(1)$ $h_i > h_j$ while $(i, j)$ doesn't belong to a FAS $\Gamma$, or
$(2)$ $h_i \le h_j$ while $(i, j)$ belongs to $\Gamma$.
When each directed arc $(i, j) \in A$ satisfies the constraint $C_{ij} (h_i, h_j) = 1$,
the set of arcs with $s_{ij} = 1$ constitutes a FAS $\Gamma$, and
the set of arcs with $s_{ij} = 0$ (or $A \backslash \Gamma$) forms an acyclic directed graph,
correspondingly with no SCC.
With the language of height configuration,
we reformulate the FAS problem as:
given a finite maximal height $H$,
we assign the heights on nodes
with as many satisfied height relation as $h_i > h_j$ for any arc $(i, j)$ as possible,
and a FAS is the set of the arcs $\Gamma$ violating height relations
with a total weight
$W(\Gamma) \equiv \sum _{(i, j) \in \Gamma} w_{ij}$.

For a finite $H$,
a height configuration with satisfied height constraints on each arc with a FAS $\Gamma$
corresponds to a fragmentation of directed networks into 
multiple segments with a largest difference of heights on vertices $< H$ and without any circle.
Only in the case as $H$ is large enough,
all the cycles with arbitrary length in an arbitrary given graph
can be removed without redundant arc contribution from loops or paths with length $\ge H$,
and the MFAS problem can be defined.
Thus the finite $H$ case
provides an upper bound for the size of MFAS.

\section*{Spin glass approach of the MFAS problem}

Based on the height representation,
we can define a spin glass model for the MFAS problem
on a directed graph $D = \{V, A\}$.
For a maximal height $H$ and the reweighting parameter $x$ (inverse temperature),
we have the partition function as

\begin{equation}
Z(H; x)
 = \sum _{\underbar{h}}
 e^{- x \sum _{(i, j) \in A} w_{ij} [1 - \theta(h_i - h_j)]}
 \prod _{(i, j) \in A} C _{ij}(h_i, h_j).
\end{equation}
The partition function
sums on the contribution from all the height configuration $\underline{h}$ with a total size $H^N$,
and only those $\underline{h}$
as all arcs $(i, j)$ with $C_{ij}(h_i, h_j) = 1$
contribute to the partition function.
As an optimization problem,
we minimize the weight sum on a FAS as
$W(\underline h) \equiv \sum _{(i, j) \in A} w_{ij} [1 - \theta(h_i - h_j)]$ for a given $H$ and $x$,
and the MFAS problem is just the case with large enough $H$ and $x$.

In the framework of cavity method of the spin glass theory
\cite{
Mezard.Montanari-2009},
we further derive the belief propagation algorithm for the spin glass model.
On each directed arc $(i, j)$,
a set of four cavity messages
$\{p _{i \rightarrow ij}^{h_i},
q _{ij \rightarrow j}^{h_j},
p _{j \leftarrow ij}^{h_j},
q _{ij \leftarrow i}^{h_i}\}$
are defined with the normalization
$\sum _{h = 0}^{H - 1} \{p _{i \rightarrow ij}^{h},
q _{ij \rightarrow j}^{h},
p _{j  \leftarrow ij}^{h},
q _{ij  \leftarrow i}^{h}\} = 1$:
$p _{i \rightarrow ij}^{h_i}$ is defined as
the probability that the node $i$ having a height $h_i$ when the height constraint on $(i, j)$ is removed,
$q _{ij \rightarrow j}^{h_j}$
the probability that the node $j$ having a height $h_j$ satisfying the height constraint on $(i, j)$ when $j$ is removed,
$p _{j \leftarrow ij}^{h_j}$
the probability that the node $j$ having a height $h_j$ when the height constraint on $(i, j)$ is removed,
$q _{ij  \leftarrow i}^{h_i}$
the probability that the node $i$ having a height $h_i$ satisfying the height constraint on $(i, j)$ when $i$ is removed.
We can understand this spin glass model in a factor graph representation
as used frequently in hard satisfiability problem
\cite{
Mezard.Montanari-2009}:
any node $i$ with a height $h_i$
is a variable node,
while the height constraint $C_{ij}(h_i, h_j)$ on any arc $(i, j)$
can be considered as a function node
accounting the interaction between the two variable nodes.

The Bethe-Peierls approximation
\cite{
Mezard.Montanari-2009}
assumes a trivial correlation among the nearest neighbors of any vertex if the vertex is removed,
leading to the marginal probability as a factorization of cavity messages,
which produces exact results in tree-like structure in a sparse random graph.
The marginal $p _{ij}^{s_{ij}}$ for any arc $(i, j)$,
or the probability that it belongs to a FAS $\Gamma$,
can be expressed with the above defined cavity probabilities as

\begin{eqnarray}
\label{eq:p0}
p _{ij}^{0}
& & \propto
\sum _{h_i = 0}^{H - 1} p _{i \rightarrow ij}^{h_i} \sum _{h_i > h_j} p _{j \leftarrow ij}^{h_j}, \\
p _{ij}^{1}
& & \propto
e^{- x w_{ij}} \sum _{h_i = 0}^{H - 1} p _{i \rightarrow ij}^{h_i} \sum _{h_i \le h_j} p_{j \leftarrow ij}^{h_j}.
\end{eqnarray}
%

We have the self-consistent equations for
$\{p _{i \rightarrow ij}^{h_i},
q _{ij \rightarrow j}^{h_j},
p _{j \leftarrow ij}^{h_j},
q _{ij \leftarrow i}^{h_i} \}$ as below.

\begin{eqnarray}
\label{eq:pi_ij}
p _{i \rightarrow ij}^{h_i}
 & = & \frac {1}{z_{i \rightarrow ij}}
 \prod _{k \in \partial ^{+} i} q _{ki \rightarrow i}^{h_i} \times
 \prod _{k \in \partial ^{-} i \backslash j} q _{ik \leftarrow i}^{h_i}, \\
\label{eq:qij_j}
q _{ij \rightarrow j}^{h_j}
 & = & \frac {1}{z_{ij \rightarrow j}}
 (\sum _{h_i > h_j} p _{i \rightarrow ij}^{h_i} + e^{- x w_{ij}} \sum _{h_i \le h_j} p _{i \rightarrow ij}^{h_i}), \\
\label{eq:pj_ij}
p _{j \leftarrow ij}^{h_j}
 & = & \frac {1}{z_{j \leftarrow ij}}
 \prod _{k \in \partial ^{+} j \backslash i} q _{kj \rightarrow j}^{h_j} \times
 \prod _{k \in \partial ^{-} j} q _{jk \leftarrow j}^{h_j}, \\
\label{eq:qij_i}
q _{ij  \leftarrow i}^{h_i}
 & = & \frac {1}{z_{ij \leftarrow i}}
 (\sum _{h_i > h_j} p _{j \leftarrow ij}^{h_j} + e^{- x w_{ij}} \sum _{h_i \le h_j} p _{j \leftarrow ij}^{h_j}).
\end{eqnarray}
while $\partial ^{+} i$ is the set of all the incoming nearest neighbors (predecessors) of node $i$,
$\partial ^{-} i$ as the set of all the out-going nearest neighbors (successors) of node $i$,
$\backslash k$ as the exclusion of $k$ from a set,
$z_{i \rightarrow ij}, z_{ij \rightarrow j}, z_{j \leftarrow ij}, z_{ij \leftarrow i}$
as corresponding normalization factors.

With the converged cavity messages,
the estimated weight sum of the MFAS is $W = \sum _{(i, j) \in A} p _{ij}^{s_{ij}} w_{ij}$,
correspondingly the energy density of the spin glass model is $e = W / N$.
A more easy-to-understand quantity of MFAS
is the occupation density $w =  W / W(A)$
where $W(A) \equiv \sum _{(i, j) \in A} w_{ij}$ is the weight sum of all arcs.
Other thermodynamic quantities of the spin glass model,
such as the free energy density and the entropy density can be found
in Supplementary Information section III,
where the details of the implementation of belief-propagation algorithms
on graph ensembles and graph instances are presented.

\begin{figure*}
\begin{center}
 \includegraphics[width = 0.95 \linewidth]{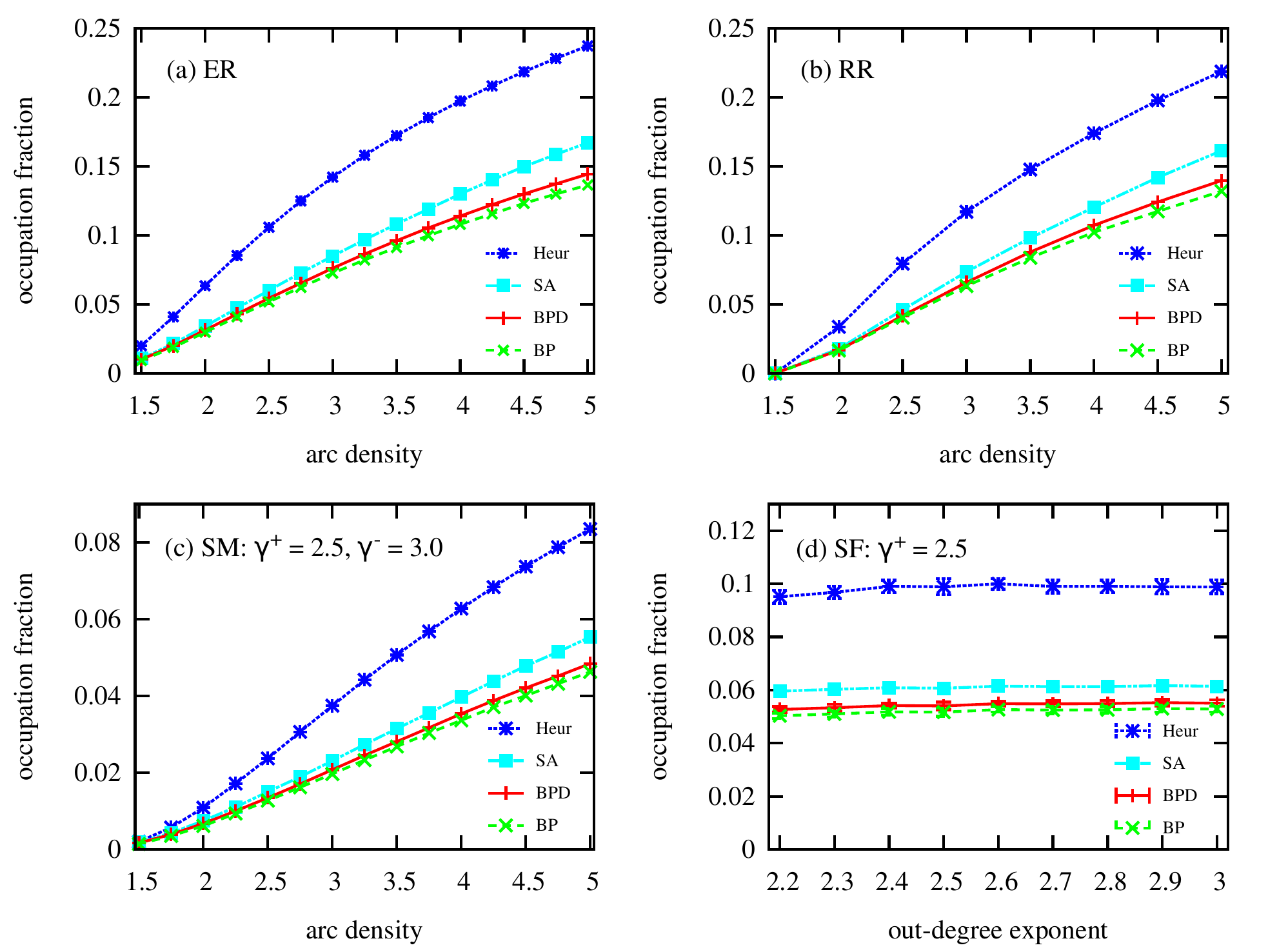}
\end{center}
\caption{
 \label{fig:fas_random}
\textbf{FAS on directed random graphs.}
We estimate the occupation density $w$ on four different models of directed random graphs
while all the arcs are assigned with uniform weight.
We apply the four methods:
(Heur) the local heuristic method based on loop-count coefficients
where at each step $0.1\%$ of the remained arcs with the largest loop-count coefficients are removed;
(SA) simulated annealing modified from the Garlinier and co-authors' method on directed feedback vertex set problem
also with the same parameters in
\cite{Garlinier.Lemamou.Bouzidi-JHeuristics-2013};
(BPD) belief propagation-guided decimation with maximal height $H = 200$ and at each decimation step
$0.5\%$ of the remained arcs with the largest marginal probabilities are removed;
(BP) belief propagation on the SCCs of the graph instances
with the maximal height $H = 200$ and the reweighting parameter $x = 50.0$.
Results from all the four methods are averaged on $40$ independently generated instances with node size $N = 10^4$.
In \textbf{(a)},
we derive FAS on
directed Erd\"{o}s-R\'{e}nyi random (ER) graphs,
while in BPD we set $x = 45.0$.
In \textbf{(b)},
we derive FAS on
directed regular random (RR) graphs,
while in BPD we set $x = 50.0$.
In \textbf{(c)},
we derive FAS on
asymptotically scale-free networks generated with static model (SM)
with in-degree exponent $\gamma ^{+} = 2.5$ and
out-degree exponent $\gamma ^{-} = 3.0$,
while in BPD we set $x = 45.0$.
In \textbf{(d)},
we derive FAS on
purely scale-free networks (SF) generated with configurational model
with an in-degree exponent $\gamma ^{+} = 2.5$ and
$k^{+}_{min} = k^{-}_{min} = 2$ and $k^{+}_{max} = k^{-}_{max} = \sqrt N$,
while in BPD we set $x = 45.0$.
In \textbf{(a)}, \textbf{(b)}, and \textbf{(c)},
the maximal variance of the results is around $1.4 \times 10^{-3}$,
and the error bars are not shown.}
\end{figure*}
\begin{table*}
\caption{
  \label{tab:real}
  \textbf{FAS on real directed networks.}
  For each real network,
  \textbf{Type} and Name
  list the its general type and name.
  $N$ and $M$
  list the numbers of its vertices and directed arcs.
  $N_{SCC}$ and $M_{SCC}$
  list the numbers of vertices and arcs in all of its SCCs.
  $N_{cl}$
  lists the number of SCC clusters.
  Heur
  lists the FAS size of a single run of the local heuristic method
  while in each step $0.1\%$ of the remained arcs with the largest loop-count coefficients are removed.
  SA
  lists the FAS size of a single run of the modified simulated annealing method based on Garlinier and co-authors' method
  with the same parameters in
  \cite{Garlinier.Lemamou.Bouzidi-JHeuristics-2013}.
  BPD
  lists the FAS size of a single run of the belief propagation-guided decimation method
  with $H = 200$ and $x = 40.0$
  while in each step $0.5\%$ of the remained arcs with the largest marginal probability on arcs are removed.}
\begin{center}
\begin{tabular}{llrrrrrrrrrr}
\hline
\textbf{Type} and Name
& $N$
& $M$
& $N_{SCC}$
& $M_{SCC}$
& $N_{cl}$
& Heur
& SA
& BPD \\
\hline
\textbf{Regulatory} \\
EGFR
& $61$
& $112$
& $61$
& $112$
& $1$
& $14$
& $9$
& $9$ \\
\textsl{S. cerevisiae}
& $688$
& $1,079$
& $3$
& $4$
& $1$
& $1$
& $1$ 
& $1$ \\
\textsl{E. coli}
& $418$
& $519$
& $0$
& $0$
& $0$
& $0$
& $0$
& $0$ \\
PPI
& $6,339$
& $34,814$
& $3,921$
& $24,164$
& $32$
& $6,536$
& $2,733$
& $2,401$\\
\hline
\textbf{Metabolic}\\
\textsl{C. elegans}
& $1,469$
& $3,447$
& $1,277$
& $3,228$
& $2$
& $939$
& $616$
& $606$ \\
\textsl{S. cerevisiae}
& $1,511$
& $3,833$
& $1,419$
& $3,719$
& $2$
& $1,167$
& $764$
& $761$ \\
\textsl{E. coli}
& $2,275$
& $5,763$
& $2,138$
& $5,568$
& $14$
& $1,736$
& $1,219$
& $1,196$ \\
\hline
\textbf{Neuronal}\\
\textsl{C. elegans}
& $297$
& $2,359$
& $243$
& $1,922$
& $3$
& $540$
& $323$
& $282$ \\
\hline
\textbf{Ecosystems}\\
Chesapeake
& $39$
& $176$
& $22$
& $60$
& $2$
& $6$
& $6$
& $7$ \\
St. Marks
& $54$
& $353$
& $33$
& $162$
& $1$
& $3$
& $3$
& $3$ \\
Florida
& $128$
& $2106$
& $103$
& $1,579$
& $1$
& $43$
& $39$
& $39$ \\
\hline
\textbf{Electric circuits}\\
s208
& $122$
& $189$
& $39$
& $47$
& $8$
& $8$ 
& $8$
& $8$  \\
s420
& $252$
& $399$
& $77$
& $93$
& $16$
& $16$
& $16$
& $16$ \\
s838
& $512$
& $819$
& $153$
& $185$
& $32$
& $32$
& $32$
& $32$ \\
\hline
\textbf{Ownership}\\
USCorp
 & $7,253$
 & $6,724$
 & $25$
 & $31$
 & $10$
 & $13$
 & $13$
 & $13$ \\
\hline
\textbf{Internet p2p} \\
 & $10,876$ 
 & $39,994$
 & $4,317$
 & $18,742$
 & $1$
 & $2,474$ 
& $1,644$
& $1,335$ \\
 Gnutella30
 & $36,682$
 & $88,328$
 & $8,490$
 & $31,706$
 & $1$
 & $4,211$
 & $2,436$
 & $1,899$  \\
 Gnutella31
 & $62,586$
 & $147,892$
 & $14,149$
 & $50,916$
 & $1$
 & $5,994$
 & $3,181$
& $2,369$ \\
\hline
\textbf{Social}\\
WiKi-Vote
 & $7,115$
 & $103,689$
 & $1,300$
 & $39,456$
 & $1$
 & $10,682$
 & $7,037$
 & $6,270$  \\
\hline
\hline
\end{tabular}
\end{center}
\end{table*}

\section*{Three methods to obtain FAS solutions}

Our method to exact FAS in network instances
based on the message-passing algorithm is the belief propagation-guided decimation method (BPD).
For a given graph instance,
BPD follows an iterative procedure consisting of three consecutive steps:
graph simplification,
message updating,
and arc decimation.
In the step of graph simplification,
we adopt Tarjan's method to exact all SCCs from the original graph or the residual graph
to make sure that the cavity messages are only defined and considered on those arcs in the SCCs.
In message updating,
we iterate messages following a randomized sequence of arcs
until the iterations converge or reach a maximal number of times.
In the decimation step,
we calculate the marginal probability $p^{1}_{ij}$ on each arc in the residual SCCs,
and remove those arcs with a given size
(for example, $0.5\%$ of the number of remained arcs)
with the largest marginals.
We repeat the above consecutive steps
until there is no SCC.
Finally, all the removed arcs resulted from the decimation procedures constitute a suboptimal FAS solution.
Details of implementation of the algorithm is in Supplementary Information section III.

As a comparison with the results based on statistical physics,
we consider another two methods,
a local heuristic method and a simulated annealing method
which both intrinsically involve no notation of heights.

The simple local heuristic method based on a local measure
inspired from
\cite{
Pardalos.Qian.Resende-JComOp-1999}:
for each arc $(i, j)$ in any SCC,
a loop-count coefficient can be defined as $k_{ij} \equiv \hat{k}^{+}_i * \hat{k}^{-}_{j}$
($\hat{k}^{+}_i$ and $\hat{k}^{-}_j $
are the number of predecessors of node $i$ and the number of successors of node $j$ in the same SCC cluster, respectively).
The removal of an arc $(i, j)$ with larger loop-count coefficient
can be assumed to have a higher probability to
remove more loops in the SCCs than those with smaller loop-count coefficients.
We construct the local heuristics as an iterative removal of a given number of arcs
with the largest loop-count coefficients
on the residual SCC structure resulted from Tarjan's method
until there is no SCC.

For the simulated annealing method,
we base our method on the Garlinier's method in
\cite{Garlinier.Lemamou.Bouzidi-JHeuristics-2013}.
The details are in Supplementary Information section IV.

\section*{FAS on directed random graphs}

We apply the local heuristic method,
the simulated annealing,
and BPD on directed random networks
with uniform weight on each arc
as in Fig.\ref{fig:fas_random}.
We test our algorithms on instances of
directed Erd\"{o}s-R\'{e}nyi random (ER) graphs
\cite{
Erdos.Renyi-PublMath-1959,
Erdos.Renyi-Hungary-1960}
with Poissonian degree distributions
and directed regular random (RR) graphs
with a uniform total degree for each node.
These directed graphs are generated by prescribing each link with a direction with equal probabilities
on corresponding undirected ER and RR graphs.
In real-world networks,
many networks are scale-free networks following power-law degree distributions
\cite{
Barabasi.Albert-Science-1999}.
We also apply the three methods on directed scale-free networks
generated with static model
\cite{
Goh.Kahng.Kim-PRL-2001}
and configurational model
\cite{
Zhou.Lipowsky-PNAS-2005}.
The details of construction of directed scale-free networks are in Supplementary Information section V.
For the four types of directed random networks,
our BPD method achieves the best result
compared with the local heuristic method
and the simulated annealing method.

\section*{FAS on directed real networks}

We further apply the local heuristic method,
the simulated annealing method,
and BPD
on $19$ real directed network instances.
See the results in Tab.\ref{tab:real}.
In the real networks,
we remove the self-loops ($(i, i)$ for a node $i$),
which are always in the FAS solution.
As our cavity messages is intrinsically defined on a factor graph
with factor nodes (height constraints on each arc) and variable nodes (vertices),
our BPD method doesn't need to be modified on those graphs
with multi-edges
(more than one arcs $(i, j)$ accounting different interactions from node $i$ to $j$)
or two-node loops
(a structure comprising of two arcs $(i,  j)$ and $(j, i)$ for a node pair $i$ and $j$).
Among the $19$ datasets,
except for one real network,
BPD achieves the smallest FAS size
to disrupt the networks,
especially on networks with moderate large size (node sizes $> 1000$).
We also find FAS on a small regulatory networks to compare with its FVS
(Supplementary Information section VI).
We further consider the case on the randomized counterparts
of the above network instances
while the connection topology is maintained yet the direction for each arc is randomized.
We find their relative sizes of SCCs with Tarjan's method
and also the FAS fraction with BPD.
See Section VI in the Supplementary Information.
On the comparison of results in the original networks and their randomized counterparts,
SCC fraction and FAS size for each type of real networks show a rather similar pattern.
This tendency provides clues to
the affect of their intrinsic evolution rules or design principles
on the structural formation of networks.

\section*{Collapse behavior of SCCs in directed random graphs and real networks}

\begin{figure*}
\begin{center}
 \includegraphics[width = 0.85 \linewidth]{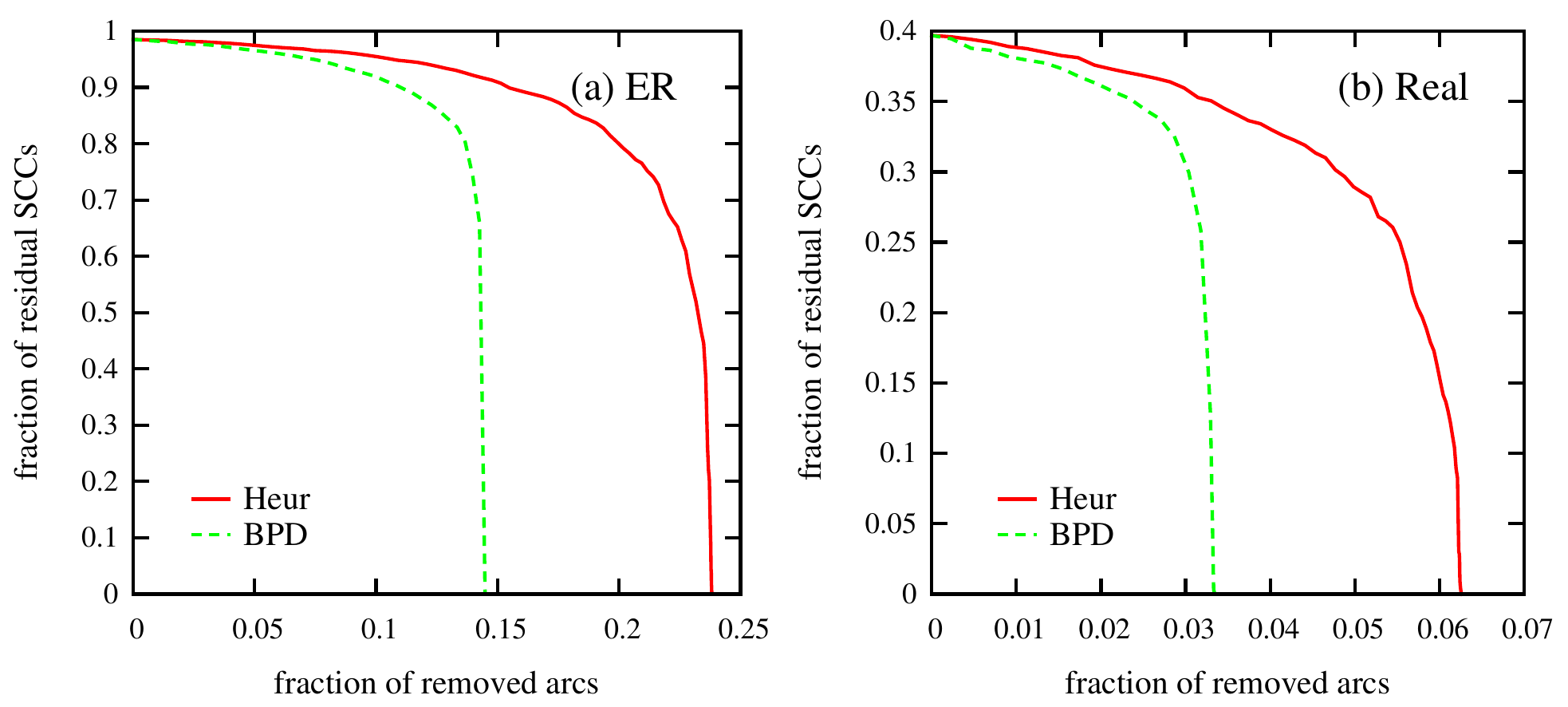}
\end{center}
\caption{
 \label{fig:col_er_real}
\textbf{SCC size in the arc removal on a random network and a real network.}
We calculate the relative sizes of all the SCCs during the arc removal process
with the local heuristics (Heur) and
the belief-propagation decimation method (BPD).
In \textbf{(a)},
we work
on a directed Erd\"{o}s-R\'enyi (ER) random graph instance with node size $N = 10^4$ with arc density $\alpha = 5.0$.
In the BPD,
the maximal height $H = 200$ and
the reweighting parameter $x = 45.0$.
In \textbf{(b)},
we work
on a real directed network instance as Gnutella04 with node size $N = 10,876$ and arc size $M = 39,994$.
In the BPD,
the maximal height $H = 200$ and
the reweighting parameter $x = 40.0$.
On both instances,
$0.5\%$ of the remained arcs with the largest loop-count coefficient are removed in Heur, and
$0.5\%$ of the remained arcs with the largest marginal probability are removed in BPD.
}
\end{figure*}

Here we consider the collapse behavior of the SCCs,
or the shrinking of the relative sizes of SCCs,
in the process of arc removal.
An example on a directed random graph instance and a real network instance
both with BPD and the local heuristic method
is in Fig.\ref{fig:col_er_real}.
In each step of removing arcs,
the BPD achieves a smaller size of SCCs.
In the last steps of arc removal,
the result with BPD experiences a more drastic jump.
It is a clear manifestation of the power of formulation as an optimization problem
taking into consideration of non-local information
over the local heuristic method considering only the local information.
As we compare the result here with the collapse behavior in undirected graphs as in
\cite{
Mugisha.Zhou-arxiv-2016},
the SCC shows a significant structural difference in a global sense than the weakly connected components (WCC).

\section*{Discussion}

The maintenance of the structural integrity and the signature dynamical behaviors
of real-world networked systems with directed interactions
are the two sides of the coin.
Here we try to ask and answer a simple yet important question since the inception of the complex network research:
how we can render a network from being 'complex' to being 'simple',
in both contexts of structure and dynamics,
in a coordinated way with hands on a smallest set of vertices or arcs
only based our knowledge of its connection topology?
We consider the optimal network disruption problem of a directed network,
or finding the minimal number of arcs to destroy \textsl{all} its strongly connected components (SCCs).
We establish the intrinsic connection of all the SCCs with
the loop structure of the directed graphs,
and further find the equivalence between the optimal network disruption problem
with the minimal feedback arc set problem (MFAS) in directed graphs,
a renown NP-hard problem in graph theory.
Equipped with the mean-field theory of spin glasses,
we define the MFAS problem into a statistical physical model
and further apply the message-passing algorithms
to extract sub-optimal disruption arc sets on network instances.

Our method has potential implications in various contexts,
such as curbing cancer growth by interrupting the interaction in gene networks in cellular processes
\cite{
Ingalls-2013},
designs of protocols with minimal effort to dysfunction a networked infrastructure system
and developing precautious measures to maintain the normal functioning of a robust system against coordinated attacks
\cite{
Cohen.Erez.benAvraha.Havlin-PRL-2001},
and the maximal dissemination of information
\cite{Kemper.Kleinberg.Tardos-ACMSIGKDD-2003}
in systems with asymmetric interactions.

Several issues of the paper are needed to further considered.
The first one is the parameter $H$ introduced in the auxiliary model.
Generally speaking, a larger $H$ leads to a better result,
and also an increasing computation time and memory.
In our result with BPD,
we remove a given fraction (for example $0.5\%$) of arcs among the SCCs with the largest marginal probabilities,
which has an approximate time complexity of $O(HM\log M)$.
Whether the integer $H$ is only an unnecessary auxiliary parameter in a better model
remains for further studies for statistical physicists and network research communities.
The second issue is that we only consider the MDAS/MFAS in the replica symmetric case
where nearly all the solutions of height configurations
are assumed to be organized in a single connected cluster.
A detailed analysis in the replica symmetry breaking case
\cite{
Mezard.Parisi-EPJB-2001}
is needed to ascertain the possible transitions of the solution configuration space
and also to devise corresponding algorithms to extract sub-optimal disruption arc sets.
The third issue
is that we consider the optimal disruption problem on simple directed networks,
a further study of the problem into the context of
multilayer networks
\cite{
Boccaletti.etal-PhysRep-2014},
which are devised in the last few years to model the intricate interactions among real-world networks,
is still needed.



\textbf{Acknowledgment}

This research is partially supported by
the National Basic Research Program of China
(grant number 2013CB932804) and by
the National Natural Science Foundations of China
(grant number 11121403 and 11225526).







\newpage

\begin{center}
Supplementary Information
\end{center}

\tableofcontents


\section{Strongly Connected Components in Directed Random Graphs}

Here we consider two algorithms to find all the strongly connected components (SCCs)
for given graph instances,
and also an analytical theory
on the relative size of the giant SCC (GSCC) in directed random graphs.

\begin{figure*}
\begin{center}
 \includegraphics[width = 0.65 \linewidth]{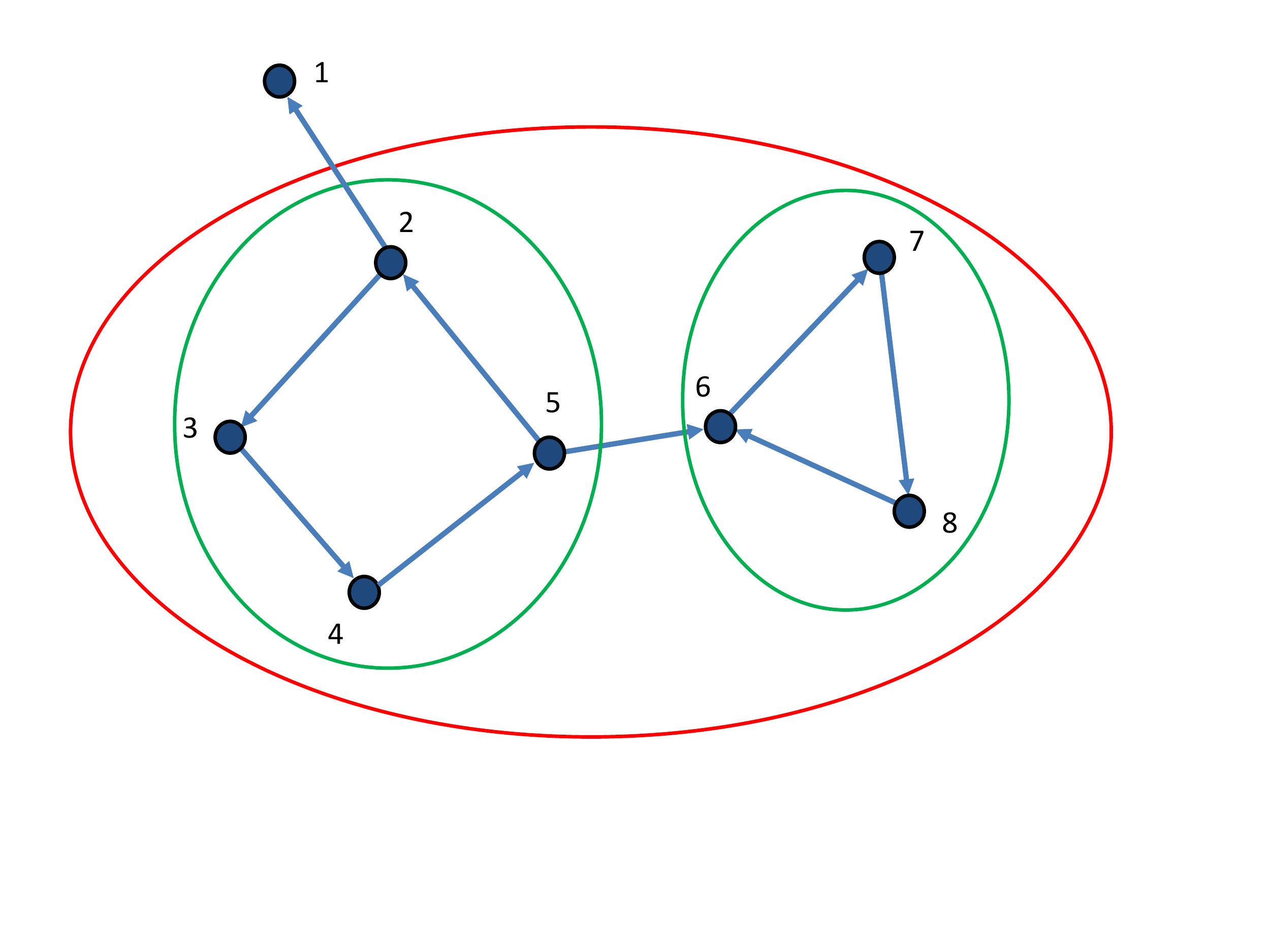}
\end{center}
\caption{
 \label{fig:scc_methods}
\textbf{Nodes and arcs in SCCs.}
We consider a simple directed graph with $8$ nodes and $9$ arcs.
With the leaf removal method,
all the nodes and arcs contained in the red circle constitute the SCC structure.
Yet with the Tarjan's method,
we can further distinguish two separate SCC clusters from the structure found by leaf removal method
which is indicated by the two green circles.}
\end{figure*}

\subsection{Leaf Removal Procedure}

As each node in a SCC cluster
belongs to certain circles,
it thus has at least one in-coming nearest neighbors and at least one out-going nearest neighbors.
We can apply a leaf removal procedure,
in which we iteratively remove all the nodes with no in-coming nearest neighbors or no out-going nearest neighbors
to reveal the SCCs.

A disadvantage of this method is that it can only reveal the nodes contained in the SCCs,
yet the collection of the arcs which constitute all the loops
is a subtle structure for the leaf removal procedure to determine.
A simple example can be in Fig.\ref{fig:scc_methods}.
As our methods based on the message-passing algorithms
involve defining and updating cavity messages on the arcs in the SCCs,
it's more important for us to find the arcs (along with the nodes) in the SCCs
rather than simply the nodes in the SCC structure for given graph instances.

\subsection{Tarjan's Method}

We adopt the Tarjan's method [31, 32]
in the decomposition of a given directed network into SCCs.
The implementation details can be found in [33].
With Tarjan's method,
both the nodes and the arcs in the graph instances can be determined.

Tarjan's method has a linear complexity as $O(N + M)$
where $N$ and $M$ are respectively the number of vertices and arcs of the given sparse graph.
Since Tarjan's method is already a much faster algorithm
than the message-passing algorithms and the local heuristic method which we will present in the next sections,
we will frequently adopt this algorithm in simulation so that we only consider algorithms on the SCCs
rather than the whole graph structure.

\subsection{Analytical Theory of the Giant Strongly Connected Component}

\begin{figure}
\begin{center}
 \includegraphics[width = 0.95\linewidth]{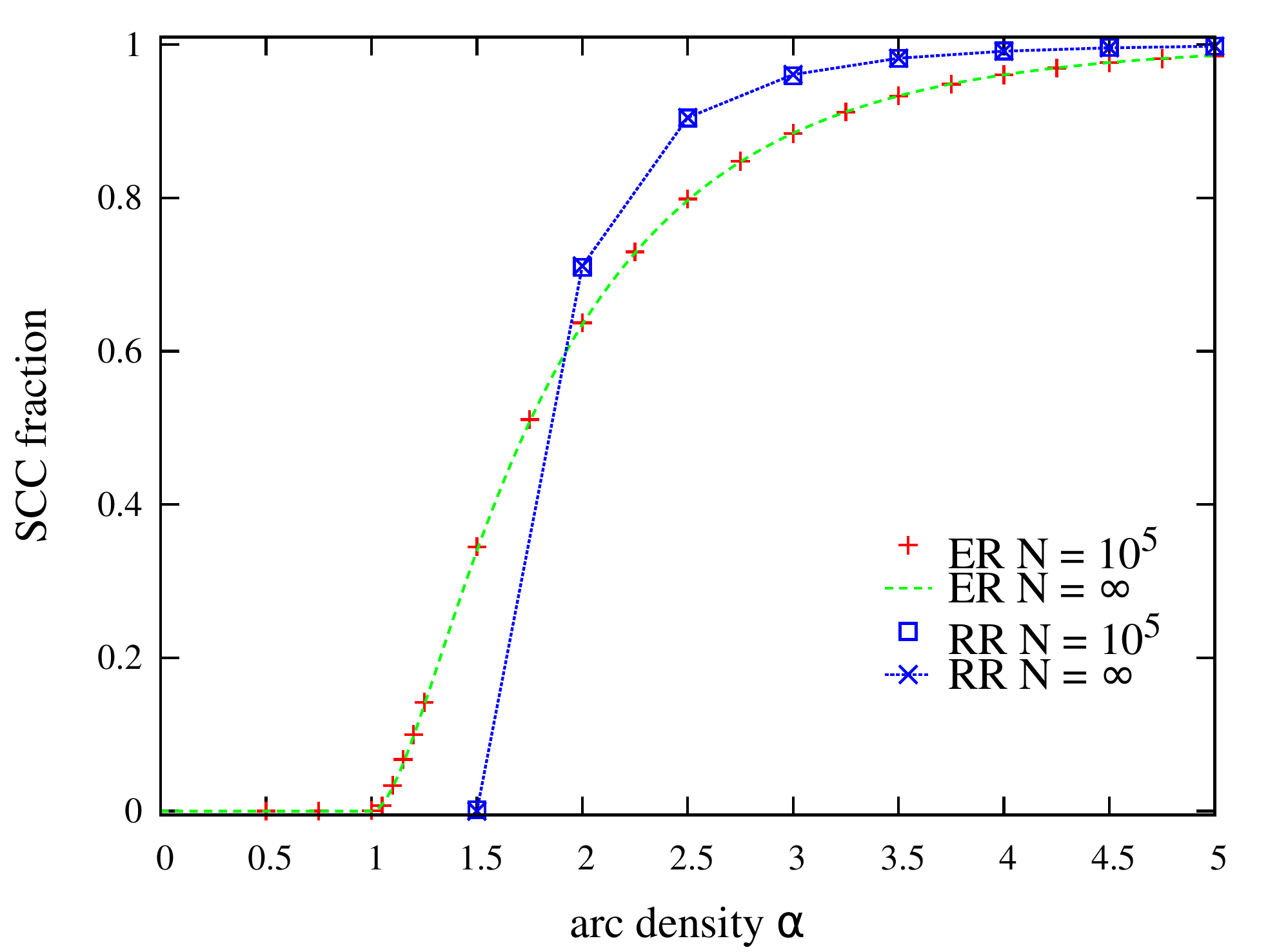}
\end{center}
\caption{
 \label{fig:scc_c}
\textbf{GSCC  and all SCCs in directed random graphs.}
The relative sizes of GSCC from mean-field theory in the case
of infinite large directed Erd\"{o}s-R\'enyi random graph (ER) and directed random regular graph (RR) are given.
We also count the fraction of nodes in all the SCCs on
single directed ER graph instances and
single directed RR graph instances with node size $N = 10^5$, respectively.}
\end{figure}

Here we reproduce some analytical theory on the percolation phenomenon of
the giant strongly connected component (GSCC) on random directed graphs [34]
to give a rough picture about the relative sizes of SCCs in directed random graphs.
We should mention that the GSCC
can only be considered as a lower bound of the size of \textsl{all} the strongly connected components.
Yet in directed random graphs,
the GSCC can be a good estimation of SCCs, as in Fig.\ref{fig:scc_c}.

For any node $i$ in a given directed graph $D = (N, A)$,
all its incoming nearest neighbors (predecessors) constitute a set $\partial ^{+} i$ with the size of in-degree $k^{+}_i$, and
all its out-going nearest neighbors (successors) constitute a set $\partial ^{-} i$ with the size of out-degree $k^{-}_i$.
The degree distribution $P(k^{+}, k^{-})$
for a random directed graph
can be defined as the probability that a randomly chosen node having
$k^{+}$ predecessors and $k^{-}$ successors.
We also define excess degree distributions.
On a randomly chosen arc $(i, j)$,
$Q^{+}(k^{+}, k^{-})$ is defined as the probability that node $j$ having $k^{+}$ predecessors and $k^{-}$ successors, and
 $Q^{-}(k^{+}, k^{-})$ is defined as the probability that node $i$ having $k^{+}$ predecessors and $k^{-}$ successors.
We can easily have

\begin{equation}
Q^{+} (k^{+}, k^{-}) = \frac {k^{+} P(k^{+}, k^{-})}{\alpha},
Q^{-} (k^{+}, k^{-}) = \frac {k^{-} P(k^{+}, k^{-})}{\alpha},
\end{equation}
where the arc density $\alpha = \sum _{k^{+}, k^{-}} k^{+} P(k^{+}, k^{-}) = \sum _{k^{+}, k^{-}} k^{-} P(k^{+}, k^{-})$.

The general line of mean-field theory of the finding GSCC
is that the GSCC can be considered as the intersection structure of the giant in-component and the giant out-component,
while the giant in-component is defined as all the nodes from which the GSCC can be reached,
and the giant out-component is defined as all the nodes can be reached from the nodes in GSCC.
For the SCC percolation,
two cavity probabilities are defined.
On a randomly directed arc $(i, j)$ in a directed graph $D = \{V, A\}$,
$x$ is defined as the probability that following node $j$ to node $i$,
node $i$ is not in the in-component;
$y$ is defined as the probability that following node $i$ to node $j$,
node $j$ is not in the out-component.
We have the self-consistent equations as

\begin{eqnarray}
x
& = &
\sum _{k^{+}, k^{-}} Q^{-}(k^{+}, k^{-}) x^{k^{+}}, \\
y
& = &
\sum _{k^{+}, k^{-}} Q^{+} (k^{+}, k^{-}) y^{k^{-}}.
\end{eqnarray}
With the stable solutions of $x$ and $y$, we can derive the normalized size of GSCC as

\begin{equation}
s = \sum _{k^{+}, k^{-}} P(k^{+}, k^{-}) (1 - x^{k^{+}}) (1 - y^{k^{-}}).
\end{equation}

In the form of generating functions, 
the general function of the directed graph with degree distribution $P(k^{+}, k^{-})$ is
$\Phi (x, y) = \sum _{k^{+}, k^{-}} P(k^{+}, k^{-}) x^{k^{+}} y^{k^{-}}$.
The percolation happens at

\begin{eqnarray}
x
 & = & \frac {\partial _{y} \Phi (x, y) |_{y = 1}}{\partial _y \Phi (1, 1)}, \\
y
 & = & \frac {\partial _{x} \Phi (x, y) |_{y = 1}}{\partial _x \Phi (1, 1)}.
\end{eqnarray}
Correspondingly, the normalized size of SCC is
$s = 1 - \Phi (x, 1) - \Phi (1, y) + \Phi (x, y)$.

\section{Equivalent Problems of FAS}

Here we consider mapping FAS on a directed graph
to its equivalent problems on a transformed graph.

\subsection{Mapping FAS to FVS}

\begin{figure*}
\begin{center}
 \includegraphics[width = 0.65 \linewidth]{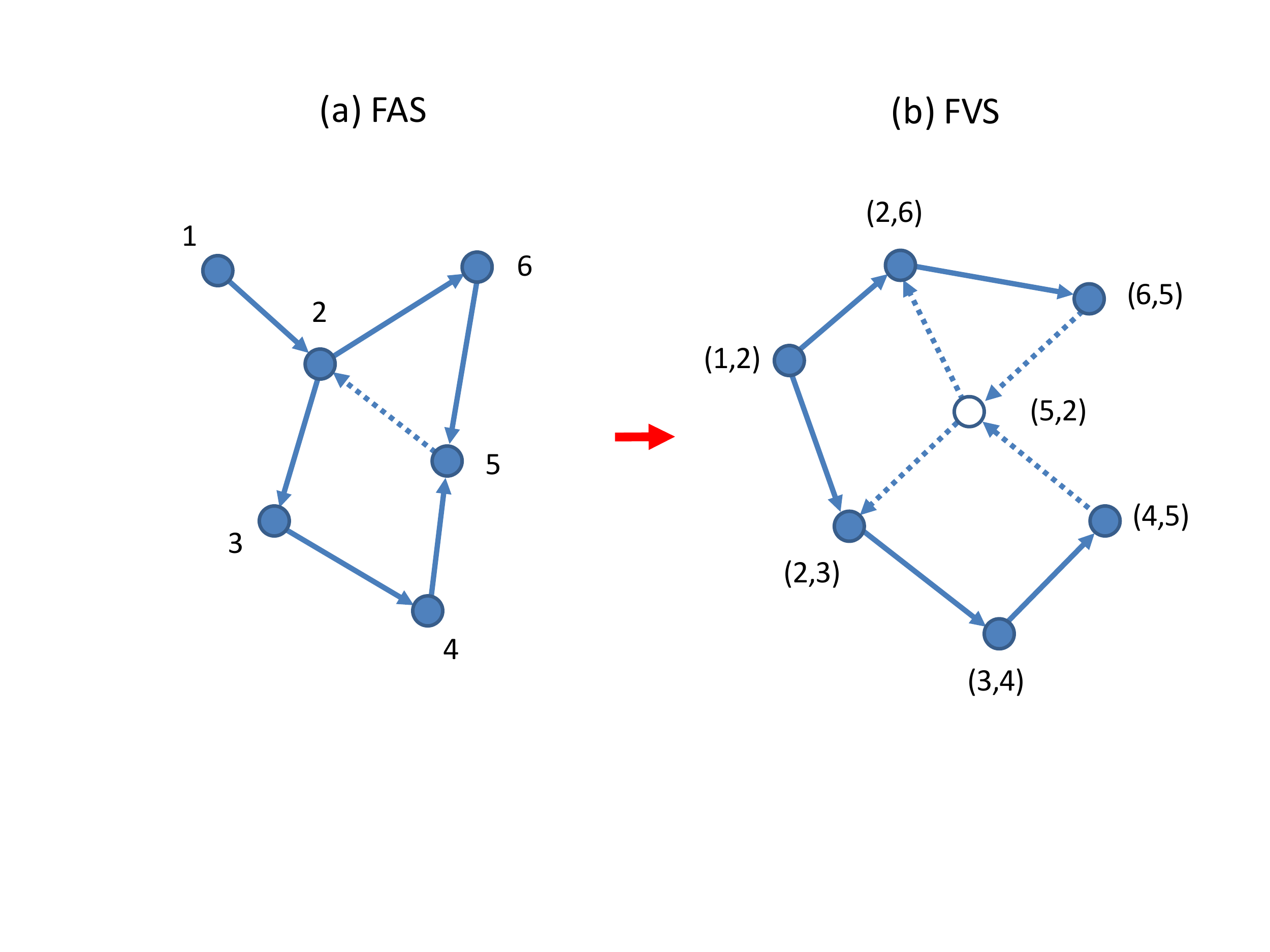}
\end{center}
\caption{
 \label{fig:fas_fvs}
\textbf{Transformation of FAS of a directed graph to FVS of its conjugate graph.}
In \textbf{(a)},
we consider a simple directed graph with $6$ nodes and $7$ arcs.
The small graph contains $2$ cycles,
as $\{2, 3, 4, 5, 2\}$ and $\{2, 6, 5, 2\}$.
The FAS is $\{(5, 2)\}$ with size $1$,
which is denoted by a dashed line.
In \textbf{(b)},
we derive the conjugate graph of the original graph,
which correspondingly contains two cycles
as $\{(2, 3), (3, 4), (4, 5), (5, 2), (2, 3)\}$ and $\{(2, 6), (6, 5), (5, 2), (2, 6)\}$.
The FVS is $\{(5, 2)\}$.
The conjugate node in FVS is denoted as empty,
and the adjacent conjugate arcs to be removed along the FVS node
are denoted in dashed lines.
}
\end{figure*}
\begin{figure*}
\begin{center}
 \includegraphics[width = 0.65 \linewidth]{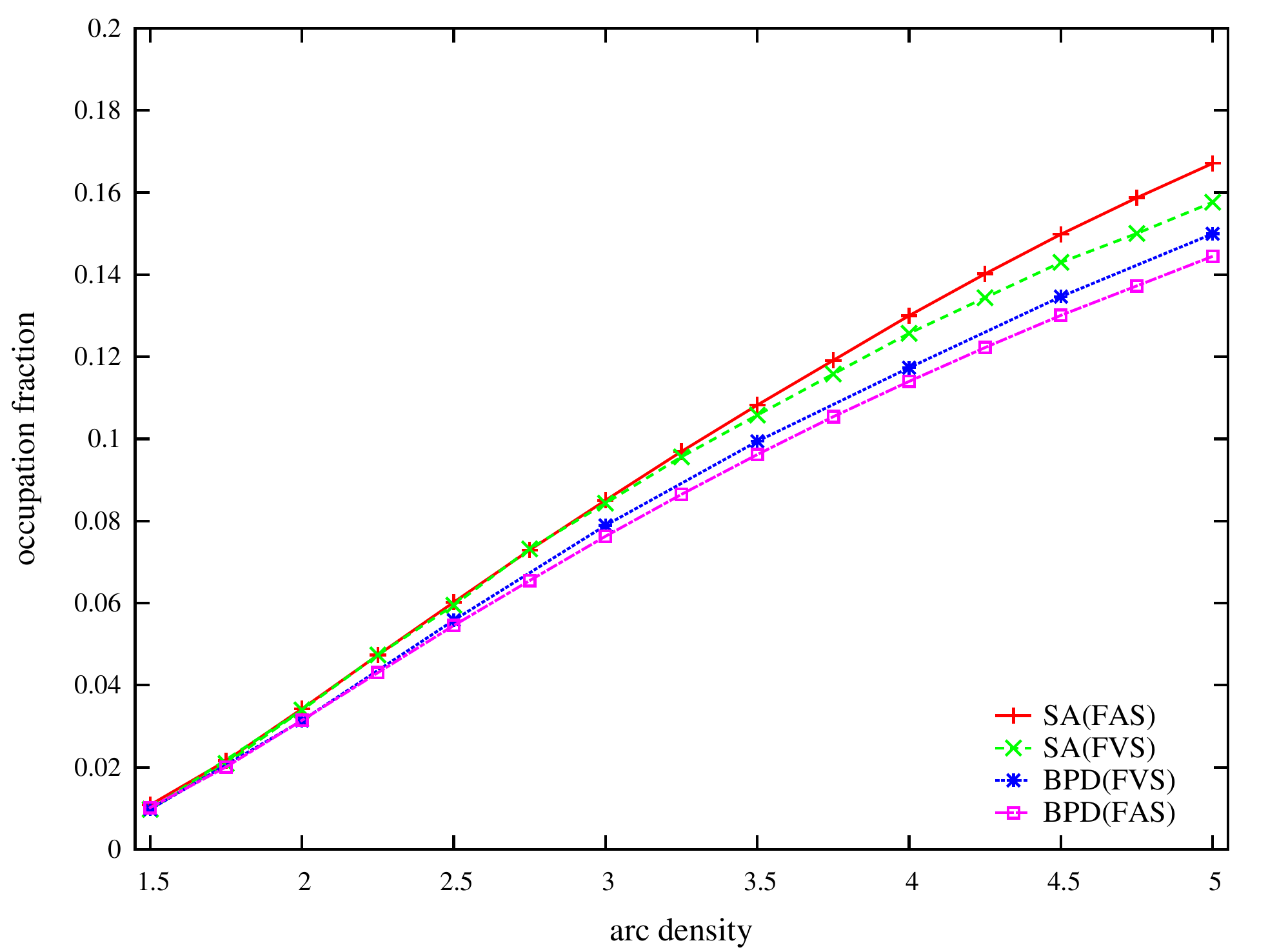}
\end{center}
\caption{
 \label{fig:fas_fvs_er}
\textbf{FAS as FVS on directed ER random graphs.}
For directed Erd\"{o}s-R\'{e}nyi random graphs with node size $N = 10^4$,
we calculate the relative FAS size with the simulated annealing method ( SA(FAS) )
and the belief propagation-guided decimation method ( BPD(FAS) )
from the main text.
We also apply the Garlinier and co-authors' simulated annealing ( SA(FVS) ) from [39]
and the belief propagation-guided decimation method ( BPD(FVS) ) from [40]
on their conjugate graph instances.
For the results on the original graphs,
we average the results on $40$ independently generated graph instances.
For the results on the conjugate graphs,
we derive the result on a single conjugate graph instance.
For the results on the BPD methods,
we set the maximal height $H = 200$,
the reweighting parameter $x = 45.0$ for the original graphs
and $x = 50.0$ for the conjugate graphs,
and at each decimation step $0.5\%$ of the remained arcs with the largest marginal probabilities
are removed.}
\end{figure*}

Here we prove that FAS in directed graphs is basically a FVS [35]
in a newly defined directed graph.

For a directed graph $D = \{V, A\}$ with a vertex set $V$ ($|V| = N$) and an arc set $A$ ($|A| = M$),
we can construct its conjugate graph $D' = \{V', A'\}$ with a vertex set $V'$ ($|V'| = N'$) and an arc set $A'$ ($|A'| = M'$)
following the two steps:
(1) arc-to-vertex correspondence:
each arc $(i, j)$ in $D$
is mapped to a vertex $(ij)$ in $D'$;
(2) connecting new vertices:
any vertices $(ij)$ and $(lm)$ in the conjugate directed graph $D'$ is connected as $((ij), (lm))$ only when $j = l$.
It's easy to see that $N' = M$ and $M' = \sum _{i \in V} k^{+}_i k^{-}_{i}$.
Following this procedure,
any directed graph has a mapped conjugate directed graph.
A feedback arc set (FAS) in a directed graph $D$ whose removal leads to an acyclic graph
is simply a feedback vertex set (FVS) in its conjugate graph $D'$.
See the example in Fig.\ref{fig:fas_fvs}.

Yet for a directed graph with moderate arc density,
the mapped conjugate graph possibly has a large expanded size of vertices and arcs.
For example,
for a directed Erd\"{o}s-R\'{e}nyi random graph instance with a node size $N$ and an arc size $M$
(correspondingly with an arc density $\alpha \equiv M / N$),
its conjugate graph has approximately a node size $\alpha N$ and an arc size $\alpha M$.
Thus the method based on message passing algorithms on conjugate graphs
needs an approximate $\alpha$ times the memory and $\alpha \log \alpha$ times the computation time
of those on the original graphs.

Apart from the consideration on the memory and time as a result of the expanded sizes of conjugate graph instances,
there is also a consideration on the results of methods on the conjugate counterparts.
See Fig.\ref{fig:fas_fvs_er}.
For directed ER random graphs,
we apply the simulated annealing method
and the belief propagation-guided decimation method
on both the original graphs in the context of FAS and
their corresponding conjugate graphs in the context of FVS.
We can see that our BPD result achieves the best results,
another advantage of devising tailored method based on statistical physics for the FAS \textsl{per se}
rather than adopting existing methods for the FVS problem on their conjugate graph instances.

\subsection{Mapping FAS to MDS}

\begin{figure*}
\begin{center}
 \includegraphics[width = 0.65 \linewidth]{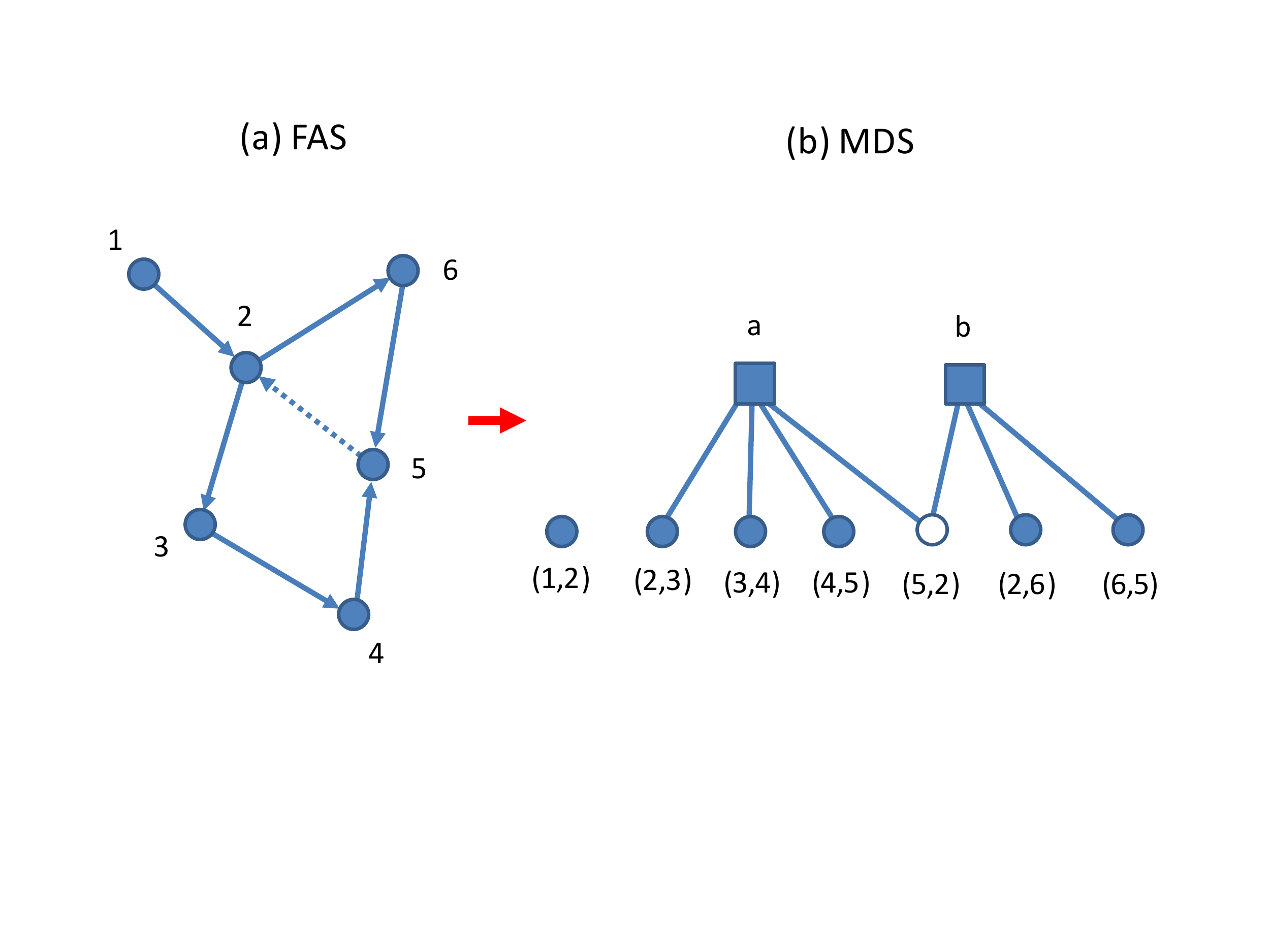}
\end{center}
\caption{
 \label{fig:fas_mds}
\textbf{Transformation of FAS of a directed graph to MDS of a bipartite graph.}
In \textbf{(a)},
we consider a simple directed graph with $6$ nodes and $7$ arcs.
The small graph contains $2$ cycles, and
the FAS is $\{(5, 2)\}$ with size $1$,
which is denoted by a dashed line.
In \textbf{(b)},
we derive a bipartite of the original graph
with $7$ A-type nodes (all the arcs in the original graph),
and $2$ B-type nodes (all the cycles in the original graph).
The MDS of the bipartite graph is simply the A-type node $(5, 2)$,
which is denoted as empty.}
\end{figure*}

Here we consider the minimal FAS problem as a minimal dominating set (MDS) problem [36-38]
of a bipartite graph.

For a given directed graph,
we first find its bipartite graph constituting of its arcs and cycles.
The two types of nodes in the bipartite graph are:
A-type node as each directed arc $(i, j)$ in the directed graph, and
B-type node as all the cycles in all the SCCs of the directed graph each of which is a set of its directed arcs.
A hyper-link is established between an A-type node $l$ and a B-type node $S$ once $l \in S$.
In the minimum dominating set problem (MDS)
in the context of a bipartite graph,
we try to find a minimal set of A-type nodes
so that each of the B-type nodes
is connected to at least one A-type nodes in this set.
See the example in Fig.\ref{fig:fas_mds}.

The hardness of this minimal dominating set (MDS) representation
originates from the very large size of B-type nodes
even in a moderate large graph cases
which renders the running time of algorithm very long.

\section{Implementation of Belief Propagation Algorithms}

Here we consider the details of implementation of the message-passing algorithms
on graph ensembles and graph instances for the DAS/FAS problem.

Before presenting the details of the belief propagation algorithm,
we list the self-consistent equations for cavity messages in belief propagation algorithm
which have been explained in the main text.

\begin{eqnarray}
\label{eq:pi_ij}
p _{i \rightarrow ij}^{h_i}
 & = & \frac {1}{z_{i \rightarrow ij}}
 \prod _{k \in \partial ^{+} i} q _{ki \rightarrow i}^{h_i} \times
 \prod _{k \in \partial ^{-} i \backslash j} q _{ik \leftarrow i}^{h_i}, \\
\label{eq:qij_j}
q _{ij \rightarrow j}^{h_j}
 & = & \frac {1}{z_{ij \rightarrow j}}
 (\sum _{h_i > h_j} p _{i \rightarrow ij}^{h_i} + e^{- x w_{ij}} \sum _{h_i \le h_j} p _{i \rightarrow ij}^{h_i}), \\
\label{eq:pj_ij}
p _{j \leftarrow ij}^{h_j}
 & = & \frac {1}{z_{j \leftarrow ij}}
 \prod _{k \in \partial ^{+} j \backslash i} q _{kj \rightarrow j}^{h_j} \times
 \prod _{k \in \partial ^{-} j} q _{jk \leftarrow j}^{h_j}, \\
\label{eq:qij_i}
q _{ij  \leftarrow i}^{h_i}
 & = & \frac {1}{z_{ij \leftarrow i}}
 (\sum _{h_i > h_j} p _{j \leftarrow ij}^{h_j} + e^{- x w_{ij}} \sum _{h_i \le h_j} p _{j \leftarrow ij}^{h_j}).
\end{eqnarray}

The marginal probability $p^{s_{ij}}_{ij}$ is

\begin{eqnarray}
\label{eq:p0}
p _{ij}^{0}
& & \propto
\sum _{h_i = 0}^{H - 1} p _{i \rightarrow ij}^{h_i} \sum _{h_i > h_j} p _{j \leftarrow ij}^{h_j}, \\
\label{eq:p1}
p _{ij}^{1}
& & \propto
e^{- x w_{ij}} \sum _{h_i = 0}^{H - 1} p _{i \rightarrow ij}^{h_i} \sum _{h_i \le h_j} p_{j \leftarrow ij}^{h_j}.
\end{eqnarray}

With the converged cavity messages, we can derive 
he free energy density
$f = ( \sum _{i \in V} f_i
+ \sum _{(i, j) \in A} f_{ij}
- \sum _{(i, j) \in A} f_{i \rightarrow ij}
- \sum _{(i, j) \in A} f_{ij \rightarrow j}) / N$, while

\begin{eqnarray}
\label{eq:fi}
f _{i}
 & = &
 - \frac {1}{x} \ln [ \sum _{h_i = 0}^{H - 1}
 \prod _{j \in \partial ^{+} i} q _{ji \rightarrow i}^{h_i} \times
 \prod _{j \in \partial ^{-} i} q _{ij \leftarrow i}^{h_i}
 ], \\
\label{eq:fij}
f _{ij}
 & = &
 - \frac {1}{x} \ln [
 \sum _{h_i = 0}^{H - 1} p _{i \rightarrow ij}^{h_i} \sum _{h_i > h_j} p _{j \leftarrow ij}^{h_j} \nonumber \\
& & 
 + e^{- x w_{ij}}
 \sum _{h_i = 0}^{H - 1} p _{i \rightarrow ij}^{h_i} \sum _{h_i \le h_j} p _{j \leftarrow ij}^{h_j}
 ], \\
\label{eq:fi_ij}
f _{i \rightarrow ij}
 & = &
 - \frac {1}{x} \ln [
 \sum _{h_i = 0}^{H - 1} p _{i \rightarrow ij}^{h_i} q _{ij \leftarrow i}^{h_i}
 ], \\
\label{eq:fij_j}
f _{ij \rightarrow j}
 & = &
 - \frac {1}{x} \ln [
 \sum _{h_j = 0}^{H - 1} q _{ij \rightarrow j}^{h_j} p _{j \leftarrow ij}^{h_j}
 ].
\end{eqnarray}
The entropy density is $s = x (e - f)$.


\subsection{BP on Random Graph Ensembles}

With the above belief propagation algorithm,
we can derive the ensemble average of FAS on random directed graphs
with population dynamics.

The population dynamics algorithm is presented as below:

\begin{description}

\item [(1)]
An array of normalized cavity messages on directed arcs are initialized randomly,
in which each element contains
$\{p_{i \rightarrow ij}^{h_i},
q_{ij \rightarrow j}^{h_j},
p_{j \leftarrow ij}^{h_j},
q_{ij \leftarrow i}^{h_i}\}$
with $h_i, h_j \in [0, H - 1]$
as $H$ is a given finite integer.
We should mention that the $i$ and $j$ in the cavity messages above are pure for notation.

\item [(2)]
The array of cavity messages are updated following equations Eqs. \ref{eq:pi_ij} - \ref{eq:qij_i}.
Each step of the message updating consists of updating two parts of messages
for each element of the message array.

\begin{description}
\item[(1)]
For the part of messages
$\{p_{i \rightarrow ij}^{h_i},
q_{ij \rightarrow j}^{h_j}\}$
with $h_i, h_j \in [0, H - 1]$:
a degree pair $\{k^{+}, k^{-}\}$
is generated following $Q^{-}(k^{+}, k^{-})$;
$k_{+}$ message elements of $q_{ij \rightarrow j}^{h_j}$ with $h_j \in [0, H - 1]$ and
$k_{-}$ message elements of $q_{ij  \leftarrow i}^{h_i}$ with $h_i \in [0, H - 1]$
are randomly selected to calculate new $p_{i \rightarrow ij}^{h_i}$ with $h_i \in [0, H - 1]$
following Eq.\ref{eq:pi_ij},
and thus $q_{ij \rightarrow j}^{h_j}$ with $h_j \in [0, H - 1]$ following Eq.\ref{eq:qij_j};
we then randomly select an element in the message array
and assign the new
$\{p_{i \rightarrow ij}^{h_i}, q_{ij \rightarrow j}^{h_j}\}$ with $h_i, h_j \in [0, H - 1]$
with the above newly calculated values correspondingly.

\item[(2)]
For the part of messages
$\{p_{j \leftarrow ij}^{h_j},
q_{ij \leftarrow i}^{h_i}\}$
with $h_i, h_j \in [0, H - 1]$:
a degree pair  $\{k^{+}, k^{-}\}$
is generated following $Q^{+}(k^{+}, k^{-})$;
$k^{+}$ message elements of $q_{ij \rightarrow j}^{h_j}$ with $h_j \in [0, H - 1]$ and
$k^{-}$ message elements of $q_{ij  \leftarrow i}^{h_i}$ with $h_i \in [0, H - 1]$
are randomly selected to calculate new
$p_{j \leftarrow ij}^{h_j}$ with $h_j \in [0, H - 1]$ following Eq.\ref{eq:pj_ij} and correspondingly
$q_{ij \leftarrow i}^{h_i}$ with $h_i \in [0, H - 1]$ following Eq.\ref{eq:qij_i},
and then assign them to
$\{p_{j \leftarrow ij}^{h_j}, q_{ij \leftarrow i}^{h_i}\}$ with $h_i, h_j \in [0, H - 1]$
correspondingly in a randomly selected element in the message array.
\end{description}

\item[(3)]
After sufficient iterations of updating messages,
we sample the messages to calculate corresponding thermodynamic quantities.

\begin{description}
\item[(3.1)]
Sampling of energy.
A sequence of pairs of message elements are randomly selected
with which we can calculate the marginal probability $p^{1}_{ij}$ with Eq.\ref{eq:p0} and \ref{eq:p1},
then we can get the energy density $\bar{e}$ of the physical model as the averaged marginals by its sample size.
The estimation of occupation density is $w = \bar{e} / \alpha$
where $\alpha$ is the arc density.

\item[(3.2)]
Sampling of free energy.
Following the degree distribution $P(k^{+}, k^{-})$,
a sequence of degree pairs $\{k^{+}, k^{-}\}$ is generated, and then
$k^{+}$ messages of $q_{ij \rightarrow j}^{h_j}$ with $h_j \in [0, H - 1]$ and
$k^{-}$ messages of $q_{ij \leftarrow i}^{h_i}$ with $h_i \in [0, H - 1]$
are randomly selected and
used as inputs to calculate the contribution of free energy $\bar{f_i}$ as in Eq.\ref{eq:fi}.
As a similar procedure as sampling energy,
the contribution of free energy $\bar{f}_{ij}$,
$\bar{f}_{i \rightarrow ij}$, and $\bar{f}_{ij \rightarrow j}$
can be calculated with Eq.\ref{eq:fij}, \ref{eq:fi_ij}, and \ref{eq:fij_j}, respectively.
Thus we can get $\bar{f}$.

\item[(3.3)]
Calculation of entropy.
With the above sampled energy density and free energy density,
we can derive the entropy density $\bar{s} \equiv x (\bar{e} - \bar{f})$.

\end{description}

\end{description}

\begin{figure*}
\begin{center}
 \includegraphics[width = 0.95 \linewidth]{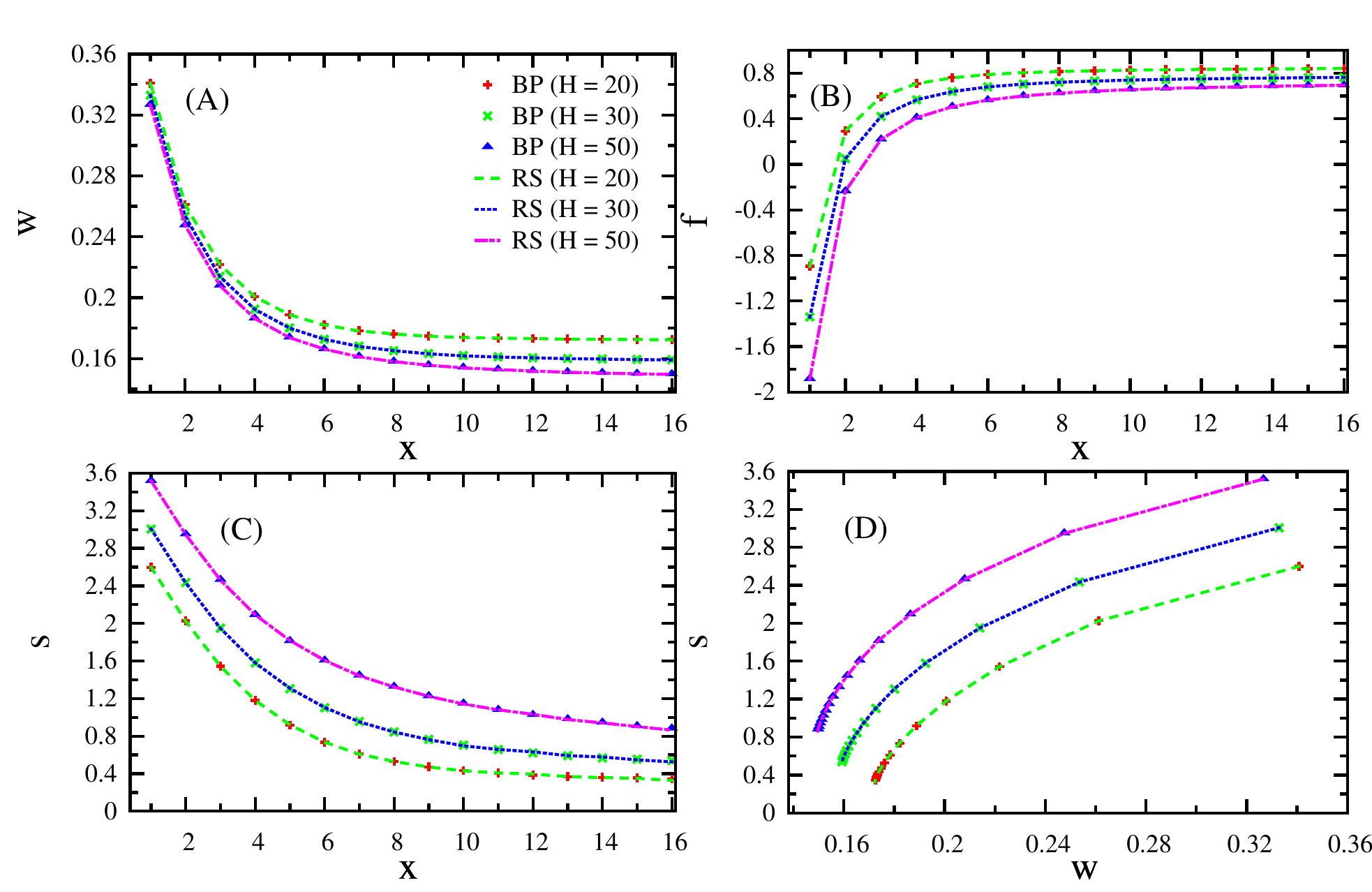}
\end{center}
\caption{
 \label{fig:fas_er_bp}
\textbf{Belief Propagation Algorithms on ER random directed graph ensembles and intances.}
We apply the belief propagation algorithm with population dynamics
on ER random directed graph ensembles with $\alpha = 5.0$ (RS, in lines)
and on ER random graphs
averaged on $40$ independently generated instances
with node size $N = 10^5$ and $\alpha = 5.0$ (BP, in signs)
with maximal height $H = 20, 30, 50$, respectively.
The occupation density $w$,
free energy density $f$, and
entropy density $s$ are calculated with different reweighting parameters $x$
in $(A)$, $(B)$, and $(C)$, respectively.
In $(D)$,
the entropy density $s$ is plotted against the occupation density $w$
as they are calculated from the same reweighting parameter $x$.
Belief propagation iterations on graph instances
converge at $x < 3.7$ for $H = 20$,
$x < 3.8$ for $H = 30$, and
$x < 3.9$ for $H = 50$.
When cavity message iterations diverge,
thermal quantities are directly calculated with messages
after a given maximal iteration number as $t_{max} = 200$.
}
\end{figure*}

An example of the result of population dynamics
can be found in Fig.\ref{fig:fas_er_bp}.

%
%


\subsection{BP on Random Graph Ensembles extrapolating to infinite $H$}

As it is clear from Fig.\ref{fig:fas_er_bp},
the occupation density gap decreases
with the same difference of maximal height with increasing $H$.
We extrapolate the occupation density $w$ on large enough reweighting parameter $x$
with increasing finite $H$ to the case of $H = \infty$.

\begin{figure}
\begin{center}
 \includegraphics[width = 0.95 \linewidth]{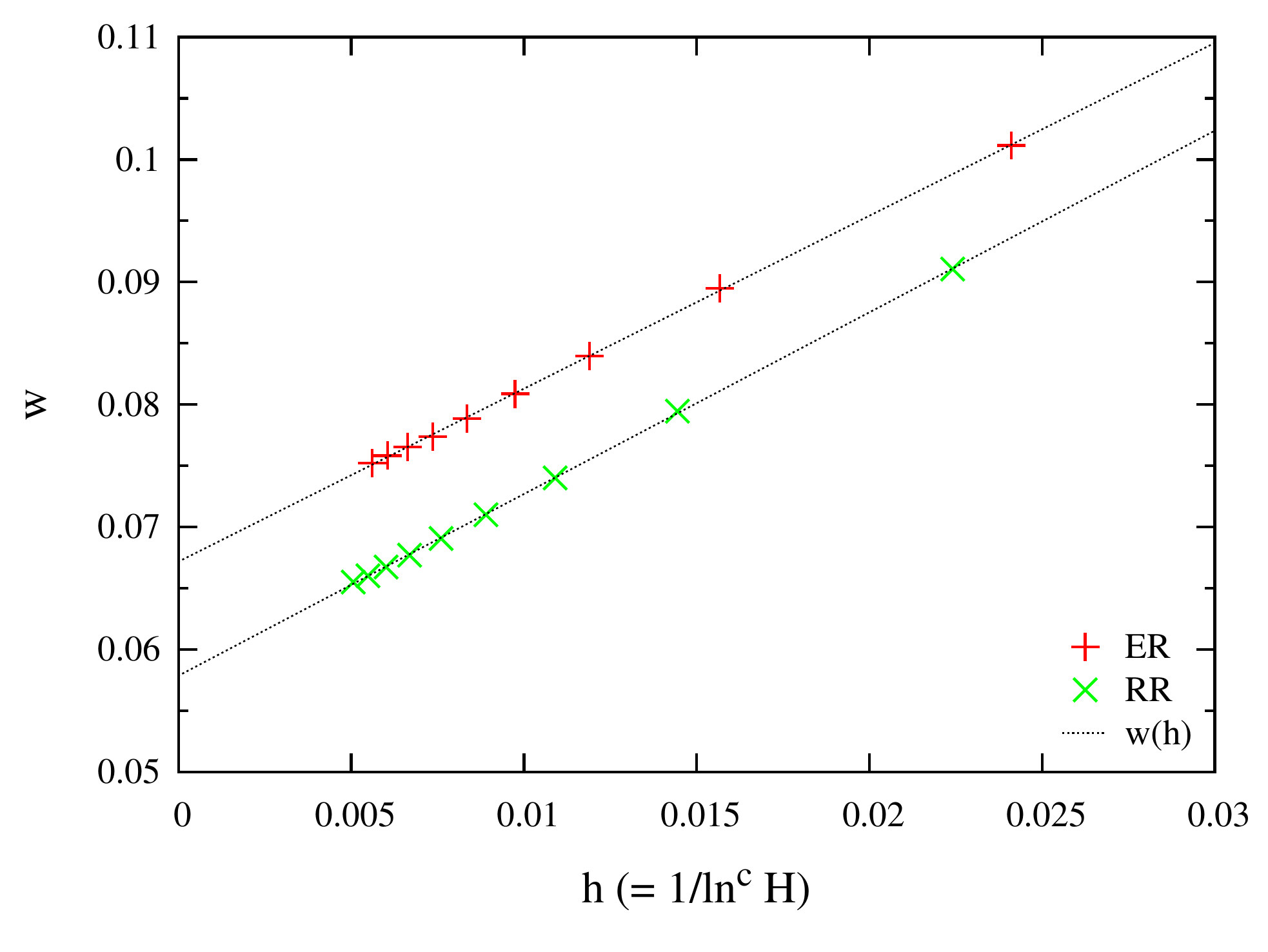}
\end{center}
\caption{
 \label{fig:fas_fitting}
\textbf{Extrapolation of occupation density to the case of $H = \infty$.}
We fit data points of the occupation density $w$
derived from the belief propagation algorithm with population dynamics
on directed ER random graph ensemble with $\alpha = 3.0$  at $x = 45$ (ER)
and on directed RR random graph ensemble with $\alpha = 3$ at $x = 50$ (RR)
with maximal heights from $H = 20$ to $100$
with the function $w(H) = a + b / \ln ^{c} (H)$.
$a$ is considered as the $w(H)$ in the case of $H = \infty$.
Here we have
$a = 0.0671751 \pm 0.0005023$ for ER random graphs, and
$a = 0.0578546 \pm 0.0003723$ for RR random graphs.}
\end{figure}

See the example in Fig.\ref{fig:fas_fitting}
for the extrapolation on ER and RR random graph ensembles with $\alpha = 3.0$ and $3$, respectively.

%

\subsection{BP on Graph Instances}

We can also apply the belief propagation algorithm
on graph instances to estimate the occupation density and other physical quantities.

\begin{description}

\item[(1)]
For a random directed graph instance $D = \{V, A\}$
with a vertex set $V$ and an arc set $A$,
on each directed arc $(i, j)$
a set of normalized cavity messages
$\{p _{i \rightarrow ij}^{h_i},
q _{ij \rightarrow j}^{h_j},
p _{j \leftarrow ij}^{h_j},
q _{ij \leftarrow i}^{h_i} \}$
with $h_i, h_j \in [0, H - 1]$
are randomly initialized.
Here $H$ is a given finite integer.

\item[(2)]
Messages on arcs are updated with a given maximal number of iterations $t_{max}$.
In each updating step,
following a randomized sequence of arcs,
new messages are calculated and assigned based on
Eqs.\ref{eq:pi_ij} - \ref{eq:qij_i}.
The message difference between two consecutive updating steps
can be defined as
the maximal absolute difference of corresponding new and old message components
$\{p _{i \rightarrow ij}^{h_i},
q _{ij \rightarrow j}^{h_j},
p _{j \leftarrow ij}^{h_j},
q _{ij \leftarrow i}^{h_i} \}$ with $h_i, h_j \in [0, H - 1]$
between updating steps $t$ and $t + 1$ with $t \ge 0$,
or $\Delta (t) = \max \{
|\delta p _{i \rightarrow ij}^{h_i} (t)|,
|\delta q _{ij \rightarrow j}^{h_j} (t)|,
|\delta p _{j \leftarrow ij}^{h_j} (t)|,
|\delta q _{ij \leftarrow i}^{h_i} (t)|\}$
with $h_i, h_j \in [0, H - 1]$.
The convergence of message updating can be easily set as
$\Delta (t) < \epsilon$
with $\epsilon$ as a small number such as $\epsilon = 10^{-8}$.

\item[(3)]
After the convergence of the messages updating or reaching the maximal iterations $t_{max}$,
with the cavity messages
we can calculate the occupation density, free energy density, and entropy density with
Eq.\ref{eq:p0} - \ref{eq:fij_j}.

\end{description}
%


\subsection{BPD on Graph Instances}

We can also apply the belief propagation-guided decimation algorithm (BPD)
on a given instance to derive a sub-optimal FAS solution.

In the definition and the updating equations of the cavity messages,
we assume the height on each vertex,
yet in the decimation method to extract a FAS from a graph instance,
we don't need actually to fix a height configuration for the graph and define the FAS as those arcs violating the height relation.
As only those arcs in the SCCs can possibly contribute to the FAS of the graph,
we adopt Tarjan's method to extract all the SCCs,
and then define and update messages only on the arcs of the SCCs,
and further use the marginals on each arc to guide the decimation of arcs in an iterative way.

For a directed graph $D = \{V, A\}$
with a vertex set $V$ and an arc set $A$,
the BPD algorithm is presented as below.

\begin{description}

\item[(1)]
For the (residual) directed graph,
we apply Tarjan's method to exact all its SCCs.
If there is no SCC, we go to Step (4).
If there is still any SCC, we go to Step (2).

\item[(2)]
We define cavity messages on each arc of the SCCs,
and run the belief-propagation algorithm to update the messages.

\item[(3)]
After the convergence of the cavity messages or reaching the maximal iterations,
the marginal probability $p^{s_{ij}}_{ij}$ on each arc $(i, j)$ is calculated.
A fraction of the arcs in SCCs (for example, $0.5\%$)
with the largest marginal probabilities are removed.
The we go to Step (1).

\item[(4)]
We count the number of the removed arcs $|\Gamma|$,
and the occupation density of the FAS is $w = |\Gamma| / |A|$.

\end{description}

\section{Simulated Annealing}

We first explain the method the simulated annealing method
by Garlinier and co-authors in [39]
which is originally designed for directed feedback vertex set problem (FVS).
Then we'll propose a modified version in the context of FAS.

In the simulated annealing procedure in [39]
we first need a convenient way to indicate the ordering of the nodes in a forest structure.
Here we can use a notational height,
which we should mention that the 'height' here is purely an intermediate way to rank nodes
and there is no range for 'heights' indicated in the algorithm.
Correspondingly,
in a tree structure,
we define a satisfied height constraint on each arc
as the predecessor node has a larger height than the successor node.
With Tarjan's method,
we only consider the nodes and the arcs in the SCC structure.
In the algorithm,
there are two complementary sets of nodes for a graph instance:
DAG as the vertices forming the forest structure, and
FVS as the vertices forming a feedback vertex set whose removal from the original graph instance leads to a forest.
The node size in FVS is considered as the energy in the simulated annealing method.
For each node in DAG, a height $h$ is defined.
For each node in FVS, two height indicators are defined,
$h_{in}$ as the minimal height of all its in-coming nearest neighbors in the DAG minus $1$, and
$h_{ou}$ as the maximal height of all its out-going nearest neighbors in the DAG plus $1$.
As an initial configuration,
all the nodes in SCCs are in the set of FVS, and no node in DAG.
Following a given cooling scheme with temperature $t = T_{0}$ and $t \rightarrow \alpha t$,
a change of the height configuration is tried:
a randomly chosen node from FVS is
randomly assigned with its $h_{in}$ or $h_{ou}$
and further moved into DAG;
if the node have no neighbor in previous DAG,
we can easily assign it with height $0$;
the new assignment may lead to the violation of height constraints on certain adjacent arcs in DAG,
then these corresponding neighbors are moved to the FVS,
thus leads to an update of the FVS size.
The new height configuration is adopted with Metropolis method.
With $maxMvt$ numbers of updating the heights of vertices in FVS and DAG
until there is no shrinkage of FVS for $maxFail$ decreasing temperatures,
we output the FVS as a suboptimal solution.

Following the general idea of the above method,
many modified versions of the simulated annealing method in the context of FAS
can be defined.
Here we consider a simple modified version.
As an initial configuration,
each node in the SCCs can be randomly assigned with a height,
for example, randomly in $[0, H_{max}]$
while $H_{max}$ can be assigned with $10$ times the node size.
The size of arcs with violated height constraints,
or the FAS,
can be considered as the energy in the simulated annealing method.
Each node is also recorded with two height indicators:
$h_{in}$ as the minimal height of all its in-coming neighbors in the SCCs minus $1$, and
$h_{ou}$ as the maximal height of all its out-going neighbors in the SCCs plus $1$.
Following a cooling scheme with temperature $t = T_{0}$ and $t \rightarrow \alpha t$,
a local change in the height configuration is tried:
for a randomly selected node,
$h_{in}$ and $h_{ou}$ are selected as its new height with an equal probability,
thus possibly results in new arcs with violating height constraints,
and correspondingly an update in FAS.
Upon the adoption of the new height configuration,
we follow the Metropolis method.
After $maxMvt$ times of updating height of vertices,
as there is no shrinkage of FAS for $maxFail$ decreasing temperature,
we output the FAS as the suboptimal solution configuration.
In the results with this modified simulated annealing method in the main text,
the parameters are set as
$T_{0} = 0.6$, $\alpha = 0.99$, $maxMvt = 5 * N$ ($N$ as the node size in graph instances), and $maxFail = 50$
just as in [39].

\section{Construction of Scale-free Networks}

Here we consider the construction of scale-free networks
generated with two kinds of methods.

\subsection{Asymptotically scale-free networks with static model}

Here we apply BPD and local heuristic method on
asymptotically scale-free (SF) networks generated with static model [41, 42].

First we consider the construction of undirected SF network instances
with degree distribution $P(k) \propto k^{- \gamma}$ with degree exponent $\gamma$,
then we consider the case of directed scale-free networks
with degree distribution $P(k^{+}) \propto (k^{+})^{- \gamma^{+}}$
and  $P(k^{-}) \propto (k^{-})^{- \gamma^{-}}$
as $\gamma ^{+}$ and $\gamma ^{-}$ are respectively the in-degree exponent and the out-degree exponent.

For the undirected scale-free networks
with node size $N$ and mean connectivity $c$
with a given degree exponent $\gamma$
we want to construct,
we can define a parameter $\xi \equiv 1 / (\gamma - 1)$.
For a graph instance with $N$ vertices with index $1, 2, .. N$ and no edges,
each node is assigned with a weight $w_i = i^{- \xi}$.
To construct an edge,
a pair of nodes not connected are chosen with respective probabilities proportional to their weights
and connected.
With this process, 
a SF network instance with $M \equiv cN/2$ edges,
can be constructed.
In the thermodynamic limit, we have the degree distribution as

\begin{eqnarray}
  P(k) & = & 
  \frac {(c (1 - \xi))^k}{\xi k!} \int _{1}^{\infty} dt e^{- c (1 - \xi) t}
  t^{k - 1 - 1 / \xi} \nonumber \\
& \equiv &
  \frac {[c(1 - \xi)/2]^{1/\xi}}{\xi} \frac {\Gamma (k - 1/\xi,
    c(1 - \xi)/2)}{\Gamma (k + 1)},
\end{eqnarray}
where $\Gamma (a)$
is the gamma function and $\Gamma (a, b)$
is the upper incomplete gamma function.
In large degree $k$, $P(k) \propto k^{- (1 + 1 / \xi)}$, or simply $P(k) \propto k^{- \gamma}$.

For the directed scale-free networks,
we follow a quite similar procedure with the undirected case.
For a directed SF network with $N$ nodes and arc density $\alpha$
with in-degree exponent $\gamma ^{+}$ and out-degree exponent $\gamma ^{-}$,
we define two parameters $\xi ^{+} \equiv 1 / (\gamma^{+} - 1)$ and $\xi ^{-} \equiv 1 / (\gamma ^{-} - 1)$.
From a graph with $N$ nodes and no arc,
each node with index $1, 2, ..., N$ are assigned with two weights as $i^{- \xi ^{+}}$ and $i ^{- \xi ^{-}}$.
In order to construct networks without in-degree and out-degree correlation,
the two sequences of nodes weights can be respectively randomized in order
thus the two weights can be decoupled from the node indices.
To construct an arc,
a node $i$ is chosen randomly proportional to its in-degree weight,
and another node $j$ is chosen randomly proportional to its out-degree weight.
If $i \neq j$ and there is no arc as $(i, j)$ nor $(j, i)$,
the arc $(i, j)$ is established in the graph.
With such procedure, $M \equiv \alpha N$ arcs are established.
Following the prove in the undirected case,
in the large $k$ limit, a SF network instance
with power-law degree distributions
with in-degree exponent $\gamma ^{+}$ and out-degree exponent $\gamma ^{-}$
can be constructed.

\subsection{Scale-free networks with configurational model}

The first method generates the scale-free networks
with power-law degree distribution based on configurational model [43].
For a scale-free network with degree exponents $\gamma ^{+}$ and $\gamma ^{-}$,
degree cut-offs are defined as
the minimal and maximal in-degrees $k^{+}_{min}$ and $k^{+}_{max}$, and
the minimal and maximal out-degrees $k^{-}_{min}$ and $k^{-}_{max}$.
An in-degree sequence and out-degree sequence are respectively constructed
based on the degree distribution
$P(k^{+}) \propto (k^{+})^{- \gamma ^{+}}$ with $k^{+}_{min} \le k \le k^{+}_{max}$ and 
$P(k^{-}) \propto (k^{-})^{- \gamma ^{-}}$ with $k^{-}_{min} \le k \le k^{-}_{max}$.
We keep the vertex sizes and the arc sizes resulted from the two degree sequences as equal,
and then we randomly connect nodes to construct arcs.

\section{FAS on Real Networks}

Here we consider a detailed comparison of the results of FAS
from a local heuristic method and BPD
on a small real network,
and then we consider the SCC and the FAS on randomized real networks,
which can provide clues to the formation of real networks.

\subsection{Real Networks}

Tab.\ref{tab:real_info}
lists some details about the $19$ real networks we use in the main text and the supplementary information.

\begin{table*}
\caption{
  \label{tab:real_info}
  \textbf{Real directed networks.}
  For each real network,
  \textbf{Type} and Name
  list the its general type and name.
  Description
  gives a further brief description for each network instance. 
  $N$ and $M$
  list the numbers of its vertices and directed arcs.}
\begin{center}
\begin{tabular}{llrrrrrrr}
\hline
\textbf{Type} and Name
& Description
& $N$
& $M$ \\
\hline
\textbf{Regulatory} \\
EGFR [44]
& Signal transduction network of EGF receptor.
& $61$
& $112$ \\
\textsl{S. cerevisiae} [45]
& Transcriptional regulatory network of \textsl{S. cerevisiae}.
& $688$
& $1,079$ \\
\textsl{E. coli} [46]
& Transcriptional regulatory network of \textsl{E. coli}.
& $418$
& $519$ \\
PPI [47]
& Protein-protein interaction network of human.
& $6,339$
& $34,814$ \\
\hline
\textbf{Metabolic}\\
\textsl{C. elegans} [48]
& Metabolic network of \textsl{C. elegans}.
& $1,469$
& $3,447$ \\
\textsl{S. cerevisiae} [48]
& Metabolic network of \textsl{S. cerevisiae}.
& $1,511$
& $3,833$ \\
\textsl{E. coli} [48]
& Metabolic network of \textsl{E. coli}.
& $2,275$
& $5,763$ \\
\hline
\textbf{Neuronal}\\
\textsl{C. elegans} [49]
& Neuronal network of \textsl{C. elegans}.
& $297$
& $2,359$ \\
\hline
\textbf{Ecosystems}\\
Chesapeake [50]
& Ecosystem in Chesapeake Bay.
& $39$
& $176$ \\
St. Marks [51]
& Ecosystem in St. Marks River Estuary.
& $54$
& $353$ \\
Florida [52]
& Ecosystem in Florida Bay.
& $128$
& $2106$ \\
\hline
\textbf{Electric circuits}\\
s208 [45]
& Electronic sequential logic circuit.
& $122$
& $189$  \\
s420 [45]
& Same as above.
& $252$
& $399$ \\
s838 [45]
& Same as above.
& $512$
& $819$ \\
\hline
\textbf{Ownership}\\
USCorp [53]
& Ownership network of US corporations.
 & $7,253$
 & $6,724$ \\
\hline
\textbf{Internet p2p} \\
Gnutella04 [54, 55]
 & Gnutella peer-to-peer file sharing network.
 & $10,876$ 
 & $39,994$ \\
 Gnutella30 [54, 55]
 & Same as above (at different time).
 & $36,682$
 & $88,328$ \\
 Gnutella31 [54, 55]
 & Same as above (at different time).
 & $62,586$
 & $147,892$ \\
\hline
\textbf{Social}\\
WiKi-Vote [56, 57]
 & Wikipedia who-votes-on-whom network.
 & $7,115$
 & $103,689$  \\
\hline
\hline
\end{tabular}
\end{center}
\end{table*}

\subsection{FAS on a Small Real Network}

Here we consider a detailed analysis of the FAS solution of a small signal transduction network
(which we can simply named as EGFR)
adapted from Fig.7 of [63]
with node size $N = 61$ and arc size $M = 112$,
whose all nodes constitutes a single SCC.
The network has been already studied in the context of feedback vertex set problem on directed network.

See Tab.\ref{tab:EGFR}.
We apply the BPD method and local heuristic method on the same network to extract FAS solutions,
where the former finds a FAS with $9$ arcs and the latter finds a FAS with $14$ arcs.
For BPD,
$3$ out the $9$ removed arcs are not the arcs with the largest loop-count coefficients
in the residual SCC structure.
Yet a smaller FAS set from BPD can still result in an acyclic directed network.

The paper [63]
finds optimal FVSs of $5$ vertices with $36$ combinations,
among which the choices of
$\{$ErbB11, ERK1/2, ADAMS, CaM, PI4,5-P2$\}$
leads to a minimal removal of $28$ arcs during the deactivation of vertices in the FVS.
As for the FAS found by the BPD method,
only $9$ arcs are needed to be removed to render the network acyclic.
Thus with the same objective to remove all the cycles in a network,
FAS offers a choice in a more controlled way on perturbing network structure.

\begin{table*}
\caption{
  \label{tab:EGFR}
  \textbf{Results on A Small Network EGFR.}
  Two methods,
  the belief-propagation decimation (BPD) and
  local heuristic method (Heuristic) are used to find FAS solutions.
  The decimation method finds $9$ arcs in the FAS,
  and the local heuristic method finds $14$ arcs in the FAS.
  The column of Removed Arc
  lists the removed arcs in the order of the decimation process.
  $LC$
  lists the loop-count coefficient of the corresponding removed arc,
  and in BPD $LC_{max}$
  lists the largest loop-count coefficient of all the remained arcs before a decimation in the SCCs.
  $SCC$ lists the node size of the remained SCCs after the arc is removed.}
\begin{center}
\begin{tabular}{ccccc|ccc}
\hline
& BPD 
&
&
& 
& Heuristic
&
& \\
\hline
& Removed Arc
& $LC$
& $LC_{max}$
& $SCC$
& Removed Arc
& $LC$
& $SCC$ \\
\hline
\hline
& (HB-EGF, ADAMS)
& $3$
& $27$
& $57$
& (ErbB11, ErbB degradation)
& $27$
& $55$ \\
& (ErbB11, ErbB degradation)
& $27$
& $27$
& $43$
& (ErbB11, SHP1)
& $18$
& $54$ \\
& (Ras, SOS)
& $4$
& $9$
& $35$
& (ErbB11,SHP2) 
& $16$
& $53$ \\
& (CaMKII, CaM)
& $3$
& $6$ 
& $24$
& (Grb2, Shc)
& $12$
& $53$ \\
& (Rac/Cdc42, SOS)
& $4$
& $4$
& $15$
& (Ras, SOS)
& $10$
& $46$ \\
& (PI4,5-P2, PLC beta)
& $2$
& $2$
& $10$
& (cyt Ca2+, RYR)
& $6$
& $45$ \\
& (ErbB11, SHP2)
& $2$
& $2$
& $9$
& (Pi4,5-P2, PI3K(p85-p110))
& $6$
& $41$ \\
& (DAG, Pi4,5-P2)
& $1$
& $1$
& $2$
& (Pi4,5-P2, PLC beta)
& $6$
& $39$ \\
& (ErbB11, SHP1)
& $1$
& $1$
& $0$
& (Pi4,5-P2, PLC gamma)
& $4$
& $38$ \\
&
&
&
&
& (ERK1/2, MKK2)
& $4$
& $37$ \\
&
&
&
&
& (ERK1/2, MKK1)
& $4$
& $29$ \\
&
&
&
&
& (HB-EGF, ADAMS)
& $3$
& $20$ \\
&
&
&
&
& (CaMKII, CaM)
& $2$
& $7$ \\
&
&
&
&
& (phosphatidyl acid, PLD)
& $1$
& $0$ \\
\hline
\hline
\end{tabular}
\end{center}
\end{table*}

\subsection{SCC and FAS on Randomized Real Networks}

\begin{figure*}
\begin{center}
 \includegraphics[width = 0.95 \linewidth]{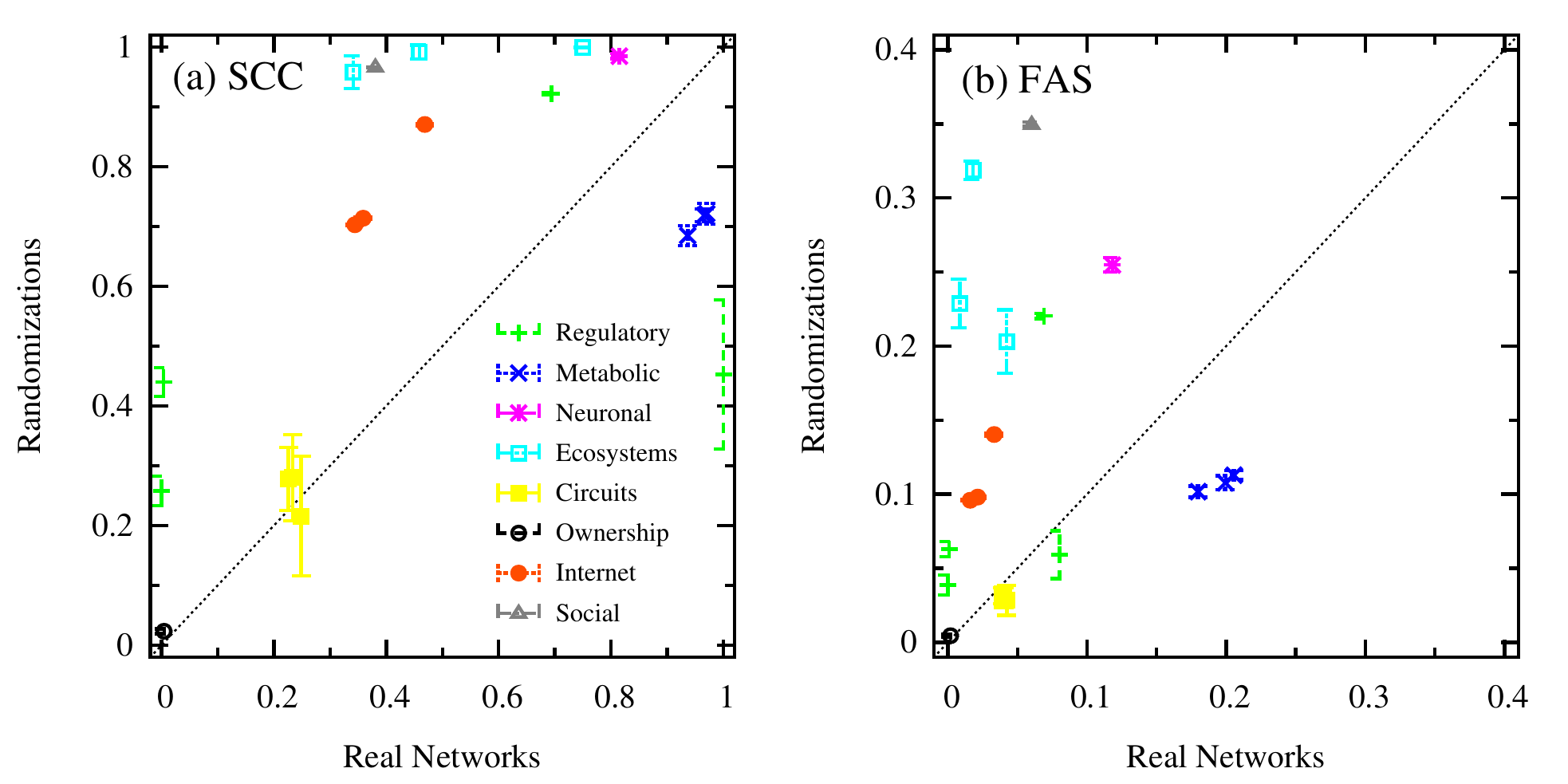}
\end{center}
\caption{
 \label{fig:real_realad}
\textbf{SCC sizes and FAS sizes on the $19$ real network instances and their counterparts with randomized arc direction.}
We calculate the relative sizes of all the SCCs in (a)
with Tarjan's method
and FAS in (b)
with belief propagation-guided decimation (BPD)
on the real network instances and their counterparts with randomized arc directions.
In BPD on real networks,
we average their results with $16$
independent initial conditions of cavity messages.
In BPD on randomized networks,
we average the results from BPD with $40$
independently generated randomized instances.
The maximal height is $H = 200$,
and the reweighting parameter is set as $x = 40.0$.
In each decimation step,
$0.5\%$ of the remained arcs with the largest marginal probabilities are removed.}
\end{figure*}

We further apply Tarjan's method and BPD method on the randomized counterparts
of the $19$ real networks with different types of interactions in the main text.
In the randomization scheme for the real network instances,
we maintain the connection topology
yet set the direction of each arc
to the original direction or its reversion with an equal probability.
See the results in Fig.\ref{fig:real_realad}.
We can see that among the $19$ real networks in different types,
most networks, especially
the networks with biological functions
($3$ of the $4$ regulatory networks,
neuronal networks, and
ecosystem networks)
and networks with social interactions
(Internet networks and the WikiVote network)
have smaller SCC sizes and FAS sizes compared with their randomized counterparts.
One type of the biological networks
(the metabolic networks)
behave quite differently from other biological networks,
as they have larger SCC sizes and FAS sizes
compared with their randomized counterparts.
The last two types of networks
(the electric circuits and the USCorp network)
which are constructed or evolve possibly following an intrinsic design principle,
show quite small difference of the SCC sizes and the FAS sizes
with those of their randomized counterparts.

From the above results,
we can draw a rather crude conclusion that:
the real world networks evolved from biological functions or social interactions (except the metabolic networks),
typically have smaller SCC sizes and FAS sizes than those in a randomized context,
thus they are easier to be disrupted by the external perturbation or forces;
the metabolic networks,
which show atypical behavior with many biological systems,
have larger size of both SCC and FAS compared with the randomized counterparts,
showing its stability against the external perturbations.

\clearpage

\newpage



\textbf{Supplementary References}

\begin{enumerate}[label={[\arabic*]}]
\setcounter{enumi}{30}

\item
{Tarjan, R. E.}
{Depth-first search and linear graph algorithms.}
{\textsl{SIAM Journal on Computing} \textbf{1}, 146-160 (1972).}

\item
{Sedgewick, R. \& Wayne, K.}
{Algorithms, fourth edition.
(Addision-Wesley, New York, 2011).}

\item
{\url{http://algs4.cs.princeton.edu/42digraph/TarjanSCC.java.html}}

\item
{Dorogovtsev, S. N., Mendes, J. F. F. \& Samukhin, A. N.}
{Giant strongly connected component of directed networks.}
{\textsl{Phys. Rev. E} \textbf{64}, 025101(R) (2001)}

\item
{Festa, P., Pardalos, P. M. \& Resende, M. G. C.}
{Feedback set problems
In \textsl{Handbook of Combinatorial Optimization, Supplement Volume A},}
{edited by Du, D.-Z. \& Pardalos, P. M. 209-258
(Springer, US, 1999)}

\item
{Haynes, T. W., Hedetniemi, S. T. \& Slater, P. J.}
{\textsl{Fundamentals of Domination in Graphs}}
{(Chapman and Hall/CRC Pure Applied Mathematics, New York, 1998).}

\item
{Zhao, J.-H., Habibulla, Y. \& Zhou, H.-J.}
{Statistical mechanics of the minimum dominating set problem.}
{\textsl{Journal of Statistical Physics} \textbf{159}, 1154-1174 (2015).}

\item
{Habibulla, Y., Zhao, J.-H. \& Zhou, H.-J.}
{The directed dominating set problem: generalized leaf removal and belief propagation,}
{Frontiers in Algorithmics, 9th International Workshop, FAW 2015}
{\textsl{Lecture Notes in Computer Science} \textbf{9130}, 78-88 (2015).}

\item
{Galinier, P., Lemamou, E. \& Bouzidi, M. W.}
{Applying local search to the feeddback vertex set problem.}
{\textsl{J. Heurisitics} \textbf{19}, 797-818 (2013)}

\item
{Zhou, H.-J.,}
{A spin glass approach to the directed feedback vertex set problem.}
{arxiv 1604.00873v1}

\item
{Goh, K.-I, Kahng, B. \& D. Kim, D.}
{Universal behavior of load distribution in scale-free networks.}
{\textsl{Phys. Rev. Lett.} \textbf{87}, 278701(2001).}

\item
{Catanzaro, M. \& Pastor-Satorras, R.}
{Analytic solution of a static scale-free network model.}
{\textsl{Eur. Phys. J. B }\textbf{44}, 241-248 (2005).}

\item
{Zhou, H.-J. \& Lipowsky, R.}
{Dynamic pattern evolution on scale-free networks.}
{\textsl{Proc. Natl. Acad. Sci. USA} \textbf{102}, 10052-10057 (2005).}


\item
{Fiedler, B., Mochizuki, A., Kurosawa, G. \& Saito, D.}
{Dynamics and control at feedback vertex sets. I:}
{Informative and determining nodes in regulatory networks.}
{\textsl{J. Dyn. Diff. Equat.} \textbf{25}, 563-604 (2013).}

\item
{Milo, R., Shen-Orr, S., Itzkovitz, S., Kashtan, N., Chklovskii, D. \& Alon, U.}
{Network motifs: simple building blocks of complex networks.}
{\textsl{Science} \textbf{298}, 824-827 (2002).}

\item
{Mangan, S. \& Alon, U.}
{Structure and function of the feed-forward loop network motif.}
{\textsl{Proc. Natl. Acad. Sci. USA} \textbf{100}, 11980-11985 (2003).}

\item
{Vinayagam, A. etal.}
{A directed protein interaction network for investigating intercellular signaling transduction.}
{\textsl{Science Signaling} \textbf{4}, rs8 (2011)}

\item
{Jeong, H., Tombor, B., Albert, R., Oltval, Z. N. \& Barab\'{a}si, A.-L.}
{The large-scale organization of metabolic networks.}
{\textsl{Nature} \textbf{407}, 651-654 (2000)}

\item
{Watts, D. J. \& Strogatz, S. H.}
{Collective dynamics of 'small-world' networks.}
{\textsl{Nature} \textbf{393}, 440-442 (1998).}

\item
{Baird, D. \& Ulanowicz, R. E.}
{The seasonal dynamics of the Chesapeake Bay ecosystem.}
{\textsl{Ecological Monographs} \textbf{59}, 329-364 (1989).}

\item
{Baird, D., Luczkovich, J. \& Christian, R. R.}
{Assessment of spatial and temporal variability in ecosystem attributes of the St Marks National Wildlife Refuge, Apalachee Bay, Florida.}
{\textsl{Estuarine, Coastal, and Shelf Science} \textbf{47}, 329-349 (1998)}

\item
{Ulanowicz, R. E., Bondavalli, C. \& Egnotovich, M. S.}
{Network Analysis of Tropic Dynamics in South Florida Ecosystem, FY 97: The Florida Bay System.}
{Ref. No. [UMCES] CBL 98-123.}
{Chesapeake Biological Laboratory, Solomons, MD 20688-0038 USA.}


\item
{Norlen, K., Lucas, G., Gebbie, M. \& Chuang, J.}
{EVA: Extraction, visualization and analysis of the telecommunications and media ownership network.}
{in Proceedings of International Telecommunications Society 14th Biennial Conference.}
{(Seoul Korea, August 2002).}

\item
{Leskovec, J., Kleinberg, J. \& Faloutsos, C.}
{Graph Evolution: Densification and Shrinking Diameters,}
{ACM Transactions on Knowledge Discovery from Data (ACM TKDD), \textbf{1} (1) (2007).}

\item
{Ripeanu, M., Foster, I. \& Iamnitchi, A.}
{Mapping the gnutella network: properties of large-scale peer-to-peer systems and implications for system design.}
{\textsl{IEEE Internet Computing} \textbf{6}, 50-57 (2002).}

\item
{Leskovec, J., Huttenlocher,  D. \& Kleinberg, J.}
{Signed networks in social media.}
{In: Proceedings of the SIGCHI Conference on Human Factors in Computing Systems, 1361-1370
(ACM, New York, 2010).}

\item
{Leskovec, J., Huttenlocher, D. \& Kleinberg, J.}
{Predicting positive and negative links in online social networks.}
{In: Proceedings of the 19th International Conference on World Wide Web, 641-650.
(ACM, New York, 2010).}

\end{enumerate}


\end{document}